**Phase separation and self-assembly in vitrimers: hierarchical morphology of molten and semi-crystalline polyethylene/dioxaborolane maleimide systems**

Ralm G. Ricarte*,†, François Tournilhac†, and Ludwik Leibler*,‡

†Matière Molle et Chimie, École Supérieure de Physique et de Chimie Industrielles de la Ville de Paris (ESPCI)–CNRS, UMR-7167, Paris Sciences et Lettres (PSL) Research University, 10 Rue Vauquelin, 75005 Paris, France and ‡UMR CNRS 7083 Gulliver, ESPCI Paris, PSL Research University, 10 Rue Vauquelin, 75005 Paris, France

Email: ralm.ricarte@espci.fr and ludwik.leibler@espci.fr

**Abstract**

Vitrimers – a class of polymer networks which are covalently crosslinked and insoluble like thermosets, but flow when heated like thermoplastics – contain dynamic links and/or crosslinks that undergo an associative exchange reaction. These dynamic crosslinks enable vitrimers to have interesting mechanical/rheological behavior, self-healing, adhesive, and shape memory properties. We demonstrate that vitrimers can self-assemble into complex meso- and nanostructures when crosslinks and backbone monomers strongly interact. Vitrimers featuring polyethylene (PE) as the backbone and dioxaborolane maleimide as the crosslinkable moiety were studied in both the molten and semi-crystalline states. We observed that PE vitrimers macroscopically phase separated into dioxaborolane maleimide rich and poor regions, and characterized the extent of phase separation by optical transmission measurements. This phase separation can explain the relatively low insoluble fractions and overall crystallinities of PE vitrimers. Using synchrotron-sourced small-angle X-ray scattering (SAXS), we discovered that PE vitrimers and their linear precursors micro-phase separated into hierarchical nanostructures. Fitting of the SAXS patterns to a scattering model strongly suggests that the nanostructures – which persist in both the melt and amorphous fraction of the semi-crystalline state – may be described as dioxaborolane maleimide rich aggregates packed in a mass fractal arrangement. These findings of hierarchical meso- and nanostructures point out that incompatibility effects between network components and resulting self-assembly must be considered for understanding behavior and the rational design of vitrimer materials.

**Introduction**

Thermoplastics and thermoplastic elastomers are soluble in good solvents and flow when heated above the glass or melting temperature. Thermosets and rubbers do not flow and are not soluble. Vitrimers, a class of polymers introduced by Leibler and collaborators in 2011, flow when heated, but remain insoluble.[1] Vitrimers are made of polymer networks which contain covalent crosslinks that undergo dynamic *associative* exchange reactions. The covalent crosslinks in a vitrimer maintain network connectivity at all times and temperatures. Unlike materials employing *dissociative* crosslinking mechanisms,[2] vitrimers cannot be completely dissolved – even in good solvents.[1,3] Associative exchange reactions permit the network topology to fluctuate and the system to flow when stress is applied, and exchange reaction kinetics control the vitrimer relaxation dynamics and viscosity.[2,4,5,6]

The initial reports of vitrimer systems focused on epoxy networks that reorganized via metal-catalyzed transesterification.[1,4,6] Today, the library of dynamic exchange reactions has expanded to include chemistries that are catalytically-controlled (olefin metathesis and transcarbonation)[7,8] or catalyst-free (transamination,[9,10,11] trans-*N*-alkylation,[12,13] reversible addition of thiols,[14] imine exchange,[15,16] addition-fragmentation chain transfer,[17,18] boronic esterification,[19,20,21] boroxine exchange,[22] and dioxaborolane metathesis).[23] Judicious choice of the exchange reaction offers an avenue for tuning vitrimer mechanical/rheological properties, self-healing and adhesion ability, and shape memory properties.[5,6,23,24,25,26] Theoretical models have been developed to clarify the relationship between the exchange reaction kinetics and resulting vitrimer material characteristics.[27,28,29,30,31,32]

While much of the vitrimer literature focuses on optimizing exchange reactions, diversifying polymer backbones, and applications,[33,34,35,36,37,38,39,40,41,42,43,44,45,46,47,48,49] less attention has been directed towards exploring incompatibility effects. This question should be particularly important in vitrimer systems, for example, in which the crosslink functional groups and network strands have very different polarizabilities. Here, we show that network element interactions can lead to hierarchical organization of vitrimers at nano- and mesoscopic length scales through micro-phase self-assembly and macroscopic phase separation. Accounting for and taking advantage of such structuring is crucial for understanding vitrimer properties and for applications.

Interactions between functional groups and the polymer backbone induce micro-phase separation in many functional polymers. For polyolefins with grafts incompatible with the backbone monomers, micro-phase separation and formation of interesting nanostructures have been reported.[50,51,52,53] In ionomer systems, where the polymer possesses a small fraction of ionic groups, strong dipole-dipole interactions between ion pairs and dielectric contrast between ion pairs and the polymer backbone drive the formation of clusters.[54,55,56,57,58] For supramolecular polymers, which have side- or end-groups engaged in non-covalent bonding, the combination of directional interactions and dispersion forces can spur the formation of aggregates or long-ranged ordered structures.[59,60,6162,63,64,65] For thermo-reversible covalent networks, van Duin and collaborators recently unveiled micro-phase separation in ethylene/propylene rubbers crosslinked with a dissociative Diels-Alder chemistry.[66]

In polymer covalent networks, macroscopic phase separation can also occur as a result of entropic effects. Indeed, when the crosslink density is sufficiently high, the presence of even good solvent may cause the network to undergo a syneresis.[67] Smallenburg *et al.* predicted that similar entropically driven phase separation into a swollen network and a dilute solution of branched clusters also occurs in dynamic vitrimer systems in the presence of solvent.[3] In bulk molten networks, chains that do not carry crosslinkable groups act as a macromolecular solvent, and phase separation could result from entropic effects – even if all components in the system are compatible.

This work shows the meso- and nanostructure of a catalyst-free vitrimer system with an enthalpically incompatible polymer backbone/crosslink moiety pair. Polyethylene (PE), a semi-crystalline polyolefin, was used as the model polymer because its strong apolarity makes it immiscible with most compounds. Following the method introduced by Röttger *et al.*,[23] the polymer was converted to a vitrimer using a two-step process. Dioxaborolane maleimide was first grafted onto the PE backbone, and subsequently crosslinked by the addition of bis-dioxaborolane (see Scheme 1). The crosslinks connecting the PE chains may rearrange via a dioxaborolane metathesis exchange reaction.[23] Multi-scale characterization by optical tools, calorimetry, and synchrotron-sourced X-ray scattering revealed that these PE vitrimers undergo complex self-assembly in both the molten and semi-crystalline state. At micron length scales, PE vitrimers macro-phase separate to form a graft-rich percolating network and graft-poor domains. At the nanoscale, enthalpic incompatibility between the PE backbone and grafts induces hierarchical micro-phase separation. The meso- and nanostructures of PE vitrimers are compared with self-assembled structures seen for other associating polymer systems, and the implications of these discoveries are discussed.

**Scheme 1.** Synthetic schemes for graft-functionalized PE and PE vitrimer.

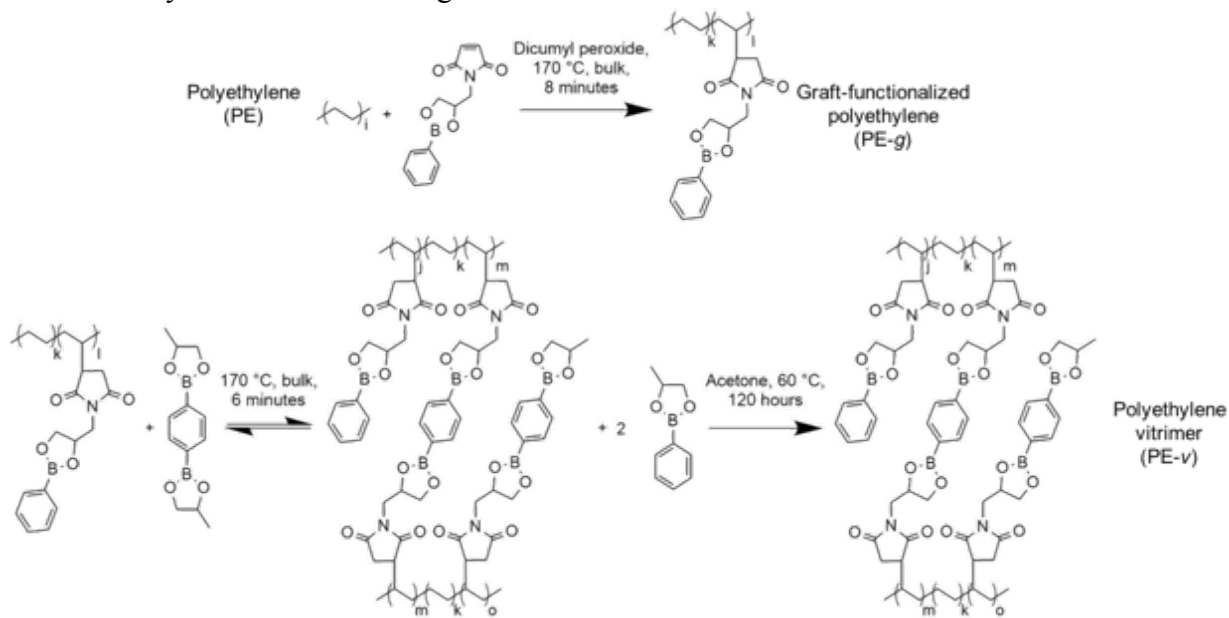

**Experimental section**

*Materials*. High-density polyethylene (PE, melt index 2.2 g/10 min at 190 °C for 2.16 kg) was purchased from Sigma-Aldrich. Tetrahydrofuran, ethanol, hexane, and xylene were purchased from Carlo Erba Reagents; furan, (±)3-amino-1,2-propanediol, and magnesium sulfate were purchased from Alfa Aesar; 1,2-propanediol and ethyl acetate were purchased from Sigma-Aldrich; phenylboronic acid was purchased from TCI Chemicals; benzene-1,4-diboronic acid was purchased from Activate Scientific.

*Synthesis of graft-functionalized polyethylene and polyethylene vitrimer.* 1-[(2-phenyl-1,3,2-dioxaborolan-4-yl)methyl]-1*H*-pyrrole-2,5-dione (*i.e.*, dioxaborolane maleimide) and 2,2'-(1,4-phenylene)-bis[4-methyl-1,3,2-dioxaborolane] (*i.e.*, bis-dioxaborolane crosslinker) were prepared according to previously reported schemes.[23] Graft-functionalized PE (PE-*g*) and PE vitrimer (PE-*v*) were synthesized using reactive mixing followed by extrusion. Functionalization and dynamic crosslinking of PE were conducted using a DSM Explore batch twin-screw extruder (capacity of 5 cm$^3$) with a co-rotating conical screw profile and a recirculation channel. The synthesis was based on the procedure reported by Röttger *et al.* (see Scheme 1).[23] Table 1 lists the sample names, targeted graft densities, measured graft densities, and measured crosslink densities. To prepare PE-*g*-4 and PE-*g*-6, PE, dioxaborolane maleimide, and dicumyl peroxide were first mixed in polypropylene bottles. The masses of PE and dicumyl peroxide were fixed to be 2.75 and 0.0014 g, respectively, while the amount of dioxaborolane maleimide was varied to target different graft densities. The mixture was fed into the extruder, which was pre-heated to a barrel temperature of 170 °C and operating at a screw speed of 100 rpm. After a residence time of 8 min, the polymer was released from the extruder. To prepare PE-*g*-8 and PE-*g*-12, the PE in the feed reagent mixture was replaced with PE-*g*-4 and PE-*g*-6, respectively. All other steps were the same as previously described. To prepare PE-*v* samples, the aforementioned functionalization step was performed, except the polymer was not released after a residence time of 8 min. Instead, at that time bis-dioxaborolane crosslinker was charged into the extruder. For PE-*v*-4-0.5, PE-*v*-6-0.5, PE-*v*-8-0.5, and PE-*v*-12-0.5, excess crosslinker was administered to fully saturate the graft moieties with crosslinker. For PE-*v*-12-0.3 and PE-*v*-12-0.2, deficient amounts of crosslinker relative to the graft moieties were fed into the extruder. After an additional residence time of 6 min, the obtained PE-*v* sample was released from the extruder. Contrary to Röttger *et al.*, each sample was washed in dry acetone under reflux conditions at 60 °C to remove residual crosslinker and small molecules.

PE and PE-*g* samples were washed for 24 hrs, while PE-*v* samples were washed for 120 hrs. Afterwards, each sample was dried under reduced pressure at 140 °C for at least 12 hrs. For comparison, peroxide crosslinked PE was also prepared (see Supporting Information).

**Table 1.** Graft and crosslink densities for PE-*g* and PE-*v*. Error bars are the standard deviation of measurements from three different pieces of the sample.

| Sample | Targeted [Graft]/[Ethylene] × $10^3$ | Measured [Graft]/[Ethylene] × $10^3$ | Measured [Crosslinker]/[Graft] |
|---|---|---|---|
| PE-*g*-4 | 4 | 3.3 ± 0.1 | 0 |
| PE-*g*-6 | 6 | 5.4 ± 0.1 | 0 |
| PE-*g*-8 | 8 | 6.9 ± 0.2 | 0 |
| PE-*g*-12 | 12 | 10.0 ± 0.6 | 0 |
| PE-*v*-4-0.5 | 4 | 4.2 ± 0.2 | 0.53 ± 0.02 |
| PE-*v*-6-0.5 | 6 | 6.2 ± 0.1 | 0.51 ± 0.01 |
| PE-*v*-8-0.5 | 8 | 7.7 ± 0.5 | 0.54 ± 0.02 |
| PE-*v*-12-0.5 | 12 | 11.6 ± 0.3 | 0.53 ± 0.02 |
| PE-*v*-12-0.3 | 12 | 10.8 ± 0.7 | 0.31 ± 0.02 |
| PE-*v*-12-0.2 | 12 | 10.7 ± 0.7 | 0.21 ± 0.02 |

*Composition quantification.* Graft- and crosslink-densities for PE-*g* and PE-*v* samples were quantified using Fourier transform infrared spectroscopy (FTIR). Measurements were made using a Bruker Tensor 37 spectrometer equipped with a Specac Goldengate attenuated total reflection cell. The PE backbone displayed $CH_2$ and $CH_3$ scissoring signals at 1470 and 1450 cm$^{-1}$, respectively. The dioxaborolane maleimide graft exhibited a C=O stretching peak at 1710 cm$^{-1}$, and aromatic C=C stretching peaks at 1600 and 1500 cm$^{-1}$. The bis-dioxaborolane crosslinker exhibited an aromatic C=C stretching signal at 1517 cm$^{-1}$.[68] The measured FTIR spectra were decomposed by fitting each characteristic peak to a skewed Gaussian function. The average graft and crosslink densities were quantified using the Beer-Lambert law. For the attenuated total reflection cell, the IR wave penetration depth into the sample ($d_p$) is nearly a linear function of the wavenumber.[69] Using this approximation and assuming molar absorption coefficients determined from solution are valid in the bulk, the PE-*g* and PE-*v* compositions were quantified using Equations 1-3:

$$\frac{C_{graft}}{C_{PE}} = \frac{A_{1710}}{A_{1470}} \frac{\varepsilon_{1470}}{\varepsilon_{1710}} \frac{\lambda_{1470}}{\lambda_{1710}} \qquad (1)$$

$$\frac{C_{crosslink}}{C_{graft}} = \frac{A_{1517}}{A_{1710}} \frac{\varepsilon_{1710}}{\varepsilon_{1517}} \frac{\lambda_{1710}}{\lambda_{1517}} \qquad (2)$$

$$\frac{C_{crosslink}}{C_{PE}} = \frac{C_{crosslink}}{C_{graft}} \times \frac{C_{graft}}{C_{PE}} \qquad (3)$$

where $C_i$ is molar concentration of component $i$, $A_j$ is the area of the peak at wavenumber $j$, $\varepsilon_j$ is the absorption coefficient of the peak at wavenumber $j$, and $\lambda_j$ is wavenumber $j$. $\varepsilon_{1710}$, $\varepsilon_{1517}$, and $\varepsilon_{1470}$ were determined through separate small-molecule solution FTIR measurements (see Figures S1 and S2).

*High-temperature size-exclusion chromatography of neat polyethylene*. To determine the molecular weight distribution of the neat PE used for the synthesis of PE-*g* and PE-*v* materials, high-temperature size-exclusion chromatography measurements were performed at the laboratory of Chemistry, Catalysis, Polymers, and Process at the University of Lyon. Measurements were performed in 1,2,4-trichlorobenzene (150 °C, 1 mL/min, stabilized by 0.2 g/L of butylated hydroxytoluene) on a Viscotek-Malvern triple detection unit. The instrument featured a refractive index detector, right-angle (90 °) and low-angle (7 °) light scattering detectors, a viscometer detector, and a combination of three separation columns from Polymer Standard Services (POLEFIN 300 mm × 8 mm; I.D porosity of $10^3$, $10^5$, and $10^6$ Å). The number- and weight-average molecular weights ($M_n$ and $M_w$) were estimated using linear polyethylene standards. The molecular weight dispersity ($Đ$) was calculated as the ratio of $M_w$ over $M_n$.

*Insoluble fraction quantification.* When placed in a good solvent, PE vitrimer leaves behind an insoluble fraction. The insoluble content of the PE-*v* samples was measured using the ISO 10147 standard procedure.[23] 0.2 g of sample was placed inside a stainless steel fine wire cage and submerged in 40 g of dried xylene. 4 g of butylated hydroxytoluene antioxidant was added to the mixture to inhibit covalent crosslinking of the PE backbone. To remove the soluble portion from the PE-*v* network, the mixture was heated under reflux conditions at 140 °C. After 18 hrs, the cage was removed from the solvent and dried under reduced pressure at 140 °C for at least 12 hrs. The mass of the remaining insoluble portion of PE-*v* was measured. The compositions of the insoluble materials were characterized by FTIR, and the insoluble fraction was determined as the mass ratio of the PE in the insoluble material to the PE in the initial vitrimer. To confirm that the insoluble fraction does not result from non-dynamic crosslinks, another extraction step was performed on the insoluble portion of PE-*v*. The procedure was similar to the previously described extraction process, but 0.5 mL of 1,2-octanediol (which breaks the bis-dioxaborolane crosslink) was also added to the refluxing mixture. The addition of the diol caused all insoluble portion of PE-*v* samples to fully dissolve in refluxing xylene, confirming the absence of non-dynamic crosslinks in the PE-*v* networks. For PE-*v*-4-0.5, the soluble portion of the vitrimer was isolated by concentrating the extraction solution under reduced pressure and subsequently precipitating the concentrated liquid in dry acetone. The precipitate was dried under reduced pressure at 140 °C for 12 hrs.

*Optical transmission measurements.* Transmission measurements were made using a custom-built setup based on previously reported experiments (see Figure S3 for the instrument schematic).[70] A piece of PE, PE-*g*, or PE-*v* sample was placed in the middle of a 1 or 0.3 mm thick Teflon™ spacer ring, sandwiched between two glass microscope slides, and put inside a Mettler FP82 hot stage. A 4 kg weight was placed on the heating stage door to compress the sample and improve optical contact between the sample and glass microscope slides. After the sample was annealed at 160 °C for 15 min, a 50 mW HeNe laser beam (wavelength 633 nm) was impinged on the sample, and a lens located downfield from the sample focused the transmitted beam onto a photodiode detector. Automated temperature control and data collection were achieved using a Mettler FP80 central processor. The transmitted intensity through the sample, $I_t$, was taken as an average of the recorded intensity over a 5 min period. $I_0$, the incident intensity, was taken to be equal to the transmitted intensity through the empty sample holder. The sample turbidity was determined using Equation S1.

*Differential scanning calorimetry (DSC)*. DSC was run using a TA Instruments Discovery DSC. 6–7 mg of sample was placed inside a Tzero aluminum pan with a standard lid. Under a nitrogen gas flow of 50 mL/min, each sample was first heated from 0–200 °C, then cooled to 0 °C, and finally reheated to 200 °C. All heating and cooling rates had a magnitude of 10 °C/min. The crystallization temperature ($T_c$) was determined from the first cooling trace, while the melting temperature ($T_m$) and enthalpy of melting ($\Delta H_m$) were measured from the second heating trace. All compounds were examined by DSC in triplicate.

*Synchrotron-sourced small-angle X-ray scattering (SAXS)*. Samples for SAXS measurements were prepared using custom-made aluminum washers (OD: 13 mm, ID: 4 mm, and thickness: 1 mm). First, one side of the washer was sealed with 25 μm thick Kapton film. 100 μm thick Kapton film and Kapton adhesive tape were placed on the outer edges of the thinner Kapton film to provide mechanical support and adhere the film, respectively. 30 mg of PE, PE-*g*, PE-*v*, or insoluble portion of PE-*v* sample were placed inside the washer opening and compression molded at 150 °C. SAXS samples were annealed under reduced pressure for 12 hrs at 140 °C to remove air bubbles. Excess sample was removed using a razor blade, and the open side of the washer was sealed using another layer of 25 μm thick Kapton film. X-ray scattering measurements were conducted at beamline SWING at the SOLEIL Synchrotron Source (Saint-Aubin, FR; range of wavevector $q$: 0.02–2 nm$^{-1}$), DND-CAT beamline 5-ID-D at the Advanced Photon Source in Argonne National Laboratory (Argonne, IL; $q$-range: 0.02–32 nm$^{-1}$), and beamline ID-02 at the European Synchrotron Research Facility (Grenoble, FR; $q$-range: 0.002–37 nm$^{-1}$). For experiments performed at ID-02, simultaneous SAXS and wide-angle X-ray scattering (WAXS) measurements were made at a sample-to-detector distance of 1 m, while only SAXS measurements were made at 3 and 31 m. The X-ray wavelength used at all sample-to-detector distances was 0. 997 Å. 2D SAXS and WAXS patterns were collected by Rayonix MX170-HS and LX170 CCD detectors, respectively. Samples were mounted inside a Mettler hot stage purged with dry nitrogen gas. At each sample-to-detector distance, SAXS patterns were first collected after annealing the sample at 160 °C for 5 min, and then after cooling the sample to 40 °C at a rate of -10 °C/min. The 2D SAXS patterns were integrated to create 1D intensity (arbitrary units) vs. wavevector $q$ (nm$^{-1}$) patterns. SAXS patterns at the different sample-to-detector distances were merged using the SAXSutilities software package,[71] and subsequently analyzed using custom-made MATLAB (version R2017b) scripts. The background removal procedure and experimental details for the DND-CAT and SWING beamline measurements are detailed in the Supporting Information.

**Results**

*Composition of dioxaborolane maleimide-grafted PE and PE vitrimer.* The samples featured in this work are labeled as PE-*g*-*α* and PE-*v*-*α*-*β*, where *α* is the targeted molar ratio multiplied by 1000 of the dioxaborolane maleimide graft to the ethylene monomer, and *β* is the targeted molar ratio of the bis-dioxaborolane crosslinker to the graft. Actual values, somewhat different from the targets, are gathered in Table 1. The graft and crosslink density of the samples were quantified using FTIR spectroscopy. Figure 1 displays FTIR spectra of PE, PE-*g*-4, and PE-*v*-4-0.5. The region between wavenumbers 1750–1420 cm$^{-1}$ is isolated for quantification because it features characteristic signals. The neat PE spectrum exhibits a CH$_2$ scissoring peak (representing the polymer backbone) and a CH$_3$ scissoring peak (reflecting the presence of branching in the chains). In addition to the polyolefin peaks, the PE-*g*-4 spectrum has a distinct C=O stretch signal at 1710 cm$^{-1}$ that originates from the maleimide segment of the graft. The phenyl ring of the graft produces two weak aromatic C=C peaks at 1600 and 1500 cm$^{-1}$. The PE-*v*-4-0.5 spectrum has the same C=O stretch signal as the precursor, but it only displays a single aromatic C=C peak at 1517 cm$^{-1}$ – this absorption is related to the crosslinker. Disappearance of the graft aromatic ring signals hints that all of the PE-*v*-4-0.5 grafts are saturated with crosslinker. In contrast, PE-*v*-12-0.3 and PE-*v*-12-0.2 exhibit all three aromatic signals (Figure S4). After decomposing the spectra to isolate the characteristic peaks (Figure S5), the graft and crosslink densities of each sample were quantified using Equations 1–3. The measured graft and crosslink densities for PE-*g* and PE-*v* samples are listed in Table 1. Vitrimers PE-*v*-4-0.5, PE-*v*-6-0.5, PE-*v*-8-0.5, and PE-*v*-12-0.5 each have a 1:2 crosslinker-to-graft molar ratio, suggesting the dioxaborolane maleimide moieties in these materials are saturated with crosslinker. Conversely, the grafts in PE-*v*-12-0.3 and PE-*v*-12-0.2 are deficient in crosslinker.

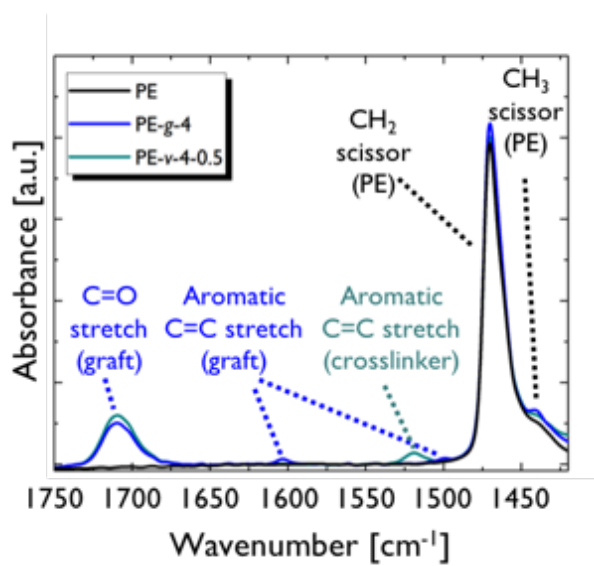

**Figure 1.** FTIR spectra of PE, PE-*g*-4, and PE-*v*-4-0.5. Characteristic signals of the PE backbone, graft, and crosslinker are highlighted.

*Macro-phase separation in the melt state.* FTIR reveals the global average composition of PE-*g* and PE-*v*. However, the distribution of grafts and crosslinks among PE chains can be non-uniform. First, the molar mass distribution of the neat PE is broad (Figure 2). This large dispersity causes the distribution of the number of grafts per PE chain to also be broad, even if the grafting is statistical (Figure S6). In this ensemble, many PE chains do not have any grafts and cannot participate in the building of a network. Second, PE and dioxaborolane maleimide are strongly interacting. It is well documented that in such a situation the grafting may be non-random.[50] More importantly, macroscopic phase separation in the molten state can occur. Figure 3 shows that a molten blend of PE and 12 wt% of dioxaborolane maleimide is turbid. Such phase separation and, more generally, ingredient incompatibility affects grafting during reactive mixing.

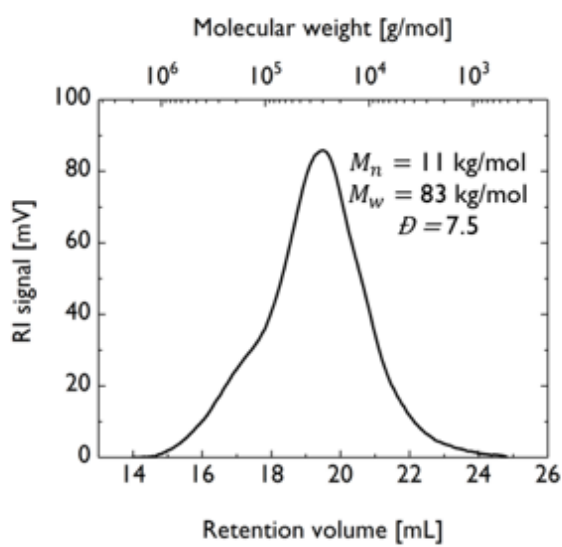

**Figure 2.** High-temperature size-exclusion chromatography trace of neat PE.

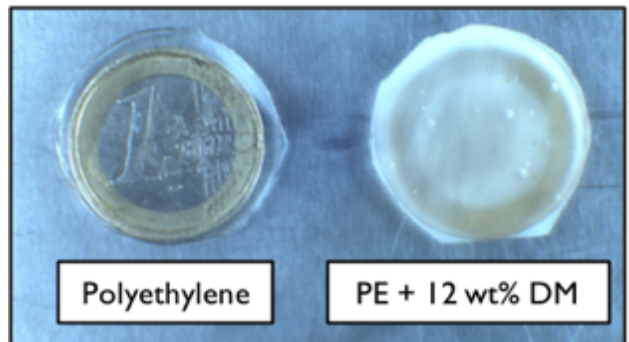

**Figure 3.** Photo of molten PE and a blend of PE + 12 wt% of dioxaborolane maleimide (DM) at 160 °C. Samples were placed on metal coins to aid visualization. PE is transparent, while the blend is turbid.

To detect and quantify possible macro-phase separation above the melting temperature of PE-*g* and PE-*v* (*i.e.*, when crystallization is turned off), we used a suite of optical tools. Visual inspection showed that unlike the neat polymer, molten PE-*g* and PE-*v* were turbid (Figure 4A), indicating bulk phase separation. Quantitatively, optical transmission measurements suggest the turbidity, which is normalized by sample thickness, arises from both bulk and surface scattering. Decreasing the sample thickness from 1 mm to 0.3 mm causes turbidity to rise, hinting that some turbidity comes from surface roughness (Figure 4B). Optical phase contrast microscopy confirms that PE-*g* and PE-*v* have rough surfaces (Figure S7). For optical turbidity measurements, the contribution from the bulk is better measured on 1 mm thick samples. At a constant thickness of 1 mm, the turbidity for both PE-*g* and PE-*v* increases as the grafting density increases (Figure 4C), thus demonstrating that grafting causes the system to macro-phase separate.

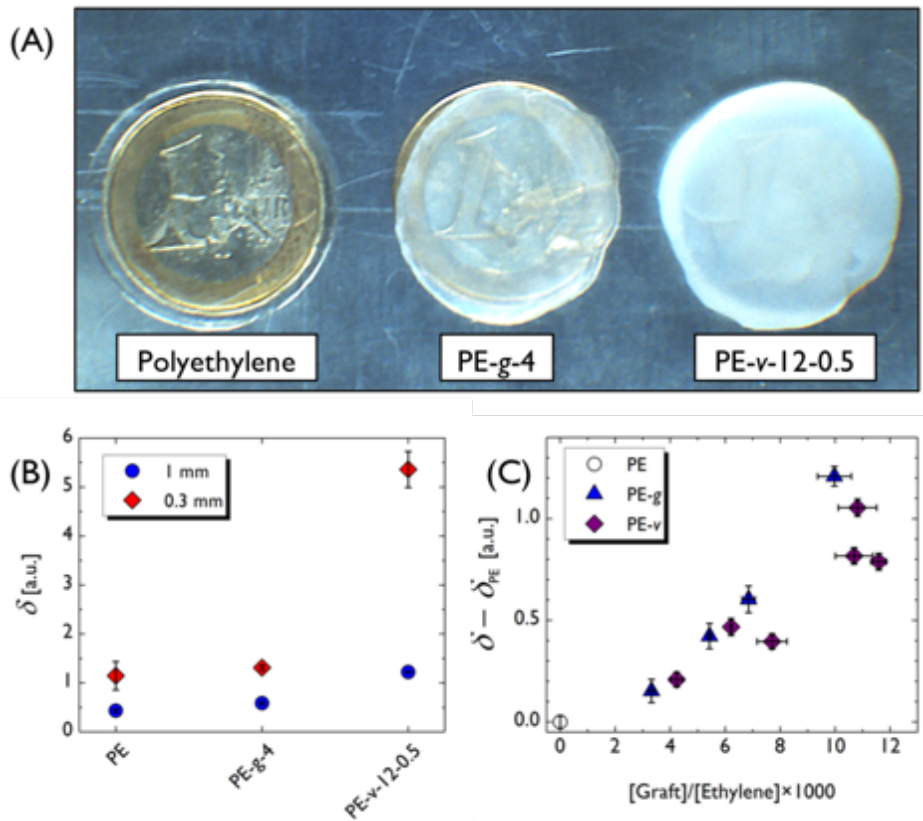

**Figure 4.** (A) Photo of molten PE, PE-*g*, and PE-*v* materials at 160 °C. Samples were placed on metal coins to aid visualization. PE is transparent, while PE-*g*-4 and PE-*v*-12-0.5 are opaque. (B) Measured turbidities ($\delta$) for the aforementioned specimens at different sample thicknesses. (C) Measured turbidities for 1 mm thick samples of PE, PE-*g*, and PE-*v*.

Macro-phase separation also causes PE-*v* materials to achieve relatively high soluble fractions. Figure 5 compares experimentally measured and expected soluble fraction values for PE-*v* materials with varying crosslink density. The solid curve, which is the estimated weight fraction of chains with 0 grafts (as calculated using Equation S2), represents the minimum possible soluble fraction for a system with statistically distributed *non-dynamic* crosslinks and the molecular weight distribution of the neat PE. The measured soluble fractions, however, are significantly higher than the lower–bound. This discrepancy arises because the soluble portion of the vitrimer not only contains chains with 0 grafts, but also chains with a small amount of grafts. As shown in Figure 6 and Table 2, the soluble portion of PE-*v*-4-0.5 shows a C=O stretch signal whose integration corresponds to a graft density of 1.2 ± 0.1 grafts per 1000 ethylene units.

Conversely, the insoluble portion of vitrimer exhibits a C=O stretch peak that correlates to 8.5 ± 0.2 grafts per 1000 ethylene units. The insoluble materials of the other PE-*v* samples are also richer in graft than their initial vitrimers (Table 2). The measured graft densities for the insoluble portion of PE-*v*-4-0.5 and the corresponding soluble portion satisfy the material balance with the initial vitrimer. Turbidity measurements and optical phase contrast microscopy of the insoluble portion of PE-*v* samples may be seen in Figure S8. Unlike the PE-*g* and PE-*v* samples, the turbidity of the insoluble portion of PE-*v* is not linearly proportional to the graft density. Furthermore, while the surface of the insoluble portion still retains a rugged topography, the feature sizes are smaller.

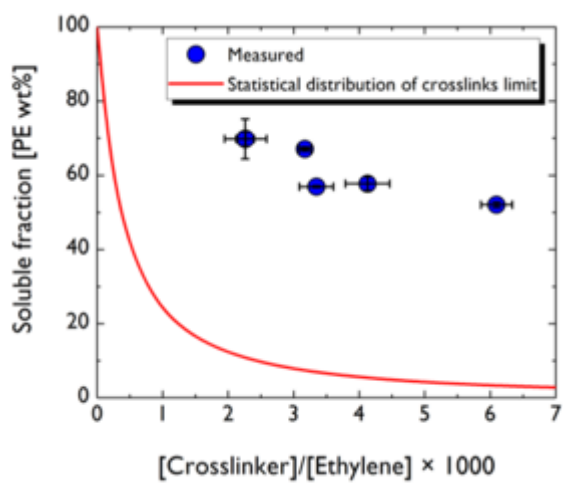

**Figure 5.** Measured soluble fractions for PE-*v* samples. The solid curve is the calculated lower-bound limit, assuming a statistical distribution of non-dynamic crosslinks. Soluble fraction values for PE-*v*-4-0.5 and PE-*v*-12-0.2 overlap at 70 wt% of PE.

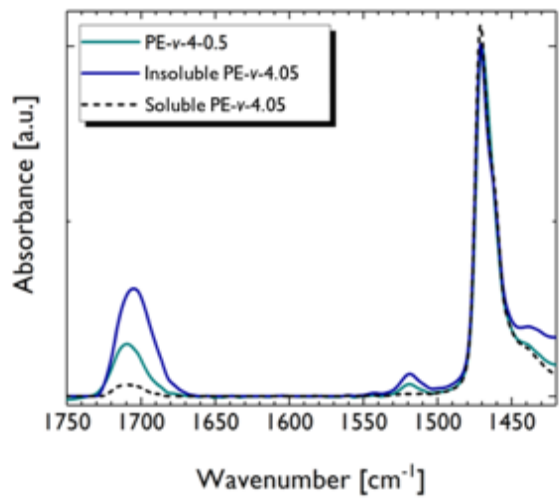

**Figure 6.** Comparison of FTIR spectra for the initial PE-*v*-4-0.5, the insoluble portion of the vitrimer, and the soluble portion of the vitrimer.

**Table 2.** Insoluble fraction, graft densities, and crosslink densities for insoluble portion of PE-*v* samples and the soluble portion of PE-*v*-4-0.5. Error bars are the standard deviation of measurements from three different pieces of the sample.

| Sample | Insoluble fraction [wt%] | [Graft]/[Ethylene] × $10^3$ | [Crosslinker]/[Graft] |
|---|---|---|---|
| Insoluble portion of PE-*v*-4-0.5 | 30 ± 5 | 8.5 ± 0.2 | 0.51 ± 0.01 |
| Insoluble portion of PE-*v*-6-0.5 | 33 ± 1 | 13.2 ± 0.1 | 0.51 ± 0.02 |
| Insoluble portion of PE-*v*-8-0.5 | 42 ± 2 | 12.9 ± 0.5 | 0.50 ± 0.01 |
| Insoluble portion of PE-*v*-12-0.5 | 48 ± 1 | 19.2 ± 0.5 | 0.51 ± 0.03 |
| Insoluble portion of PE-*v*-12-0.3 | 43 ± 1 | 22 ± 2 | 0.31 ± 0.02 |
| Insoluble portion of PE-*v*-12-0.2 | 30 ± 1 | 24 ± 3 | 0.21 ± 0.01 |
| Soluble portion of PE-*v*-4-0.5 |  | 1.2 ± 0.1 | 0.11 ± 0.01 |

As mentioned previously, the expulsion of graft-poor chains from the percolating vitrimer network is propelled by not only PE/dioxaborolane maleimide incompatibility, but also entropy.

In the case of the PE-*v* melt, the soluble portion acts as a macromolecular solvent that swells the vitrimer.

*Macro-phase separation in the semi-crystalline state.* Modification of PE by grafting, copolymerization, or crosslinking generally leads to a lower overall crystallinity fraction, though the extent of that decrease depends on the chemistry. For example, grafting of maleic anhydride onto high-density PE causes minimal changes in crystallinity, while the addition of glycidyl methacrylate creates a moderate change.[72,73] For PE ionomers, neutralization of the pendant acid groups leads to large decreases in crystallinity and may cause a low temperature melting transition to appear.[74,75,76,77] For peroxide and radiation crosslinked PE, the crystallinity decreases as the crosslink density and, vicariously, the insoluble fraction increase.[78,79,80]

For PE-*v*, macro-phase separation significantly influences the relationship among the crystallinity, crosslink density, and the insoluble fraction. Figure 7 displays the normalized crystallinity of PE, PE-*g*, PE-*v*, and insoluble portion of PE-*v* samples. (Figure S9 is the corresponding DSC traces, Figure S10 displays the non-normalized crystallinities, and Table S1 tabulates the crystallization temperature, melting point, and melting enthalpy). The normalized crystallinity of the PE-*g* samples is within error of the neat PE crystallinity, suggesting that grafting of the dioxaborolane maleimide has minimal effect on the PE crystallinity fraction. The PE-*v* data shows that the addition of crosslinker causes the crystallinity to decrease to ~ 65 wt% of PE, and removal of the soluble portion further reduces the crystallinity to 55 wt% of PE. The crystallinity of both the initial PE-*v* and insoluble portion of PE-*v* show a weak dependence on the crosslink density. At the same time, the normalized crystallinity is inversely proportional to the insoluble fraction. Similar to peroxide and radiation crosslinked PE,[79] this trend hints that the chains incorporated in the vitrimer network and those in the soluble portion of the vitrimer may crystallize separately. The dynamic crosslinks possibly slow crystallization kinetics, leading to a lower crystallinity in the graft-rich network domain. Although further studies are needed to illuminate the mechanism of crystallization in vitrimer systems, the DSC data highlight that macro-phase separation affects the crystallinity of PE-*g* and PE-*v* materials.

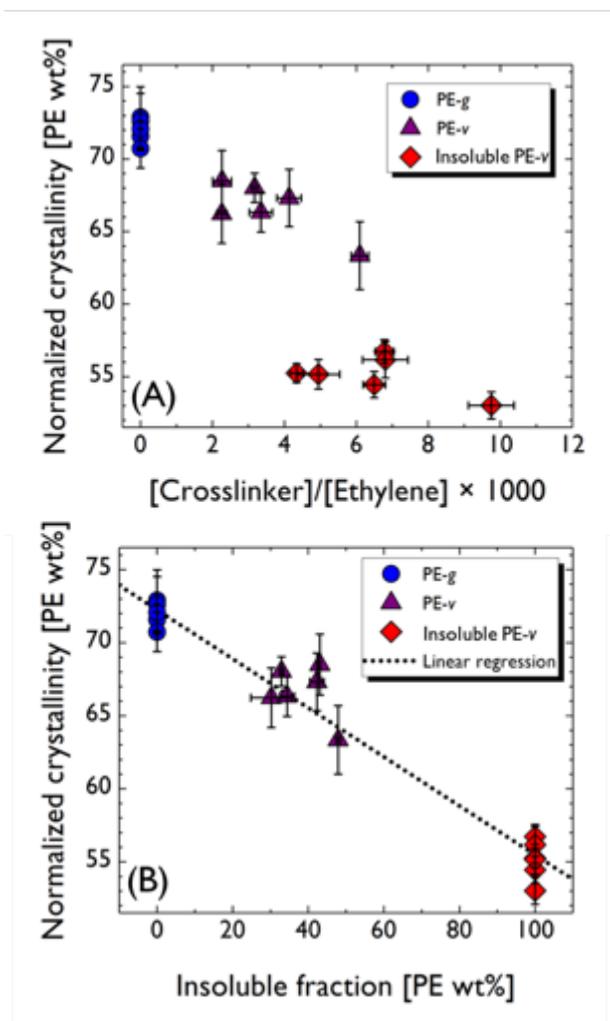

**Figure 7.** Normalized crystallinities of PE-*g*, PE-*v*, and insoluble portion of PE-*v* versus (A) crosslink density and (B) insoluble fraction.

*Micro-phase separation in the melt and semi-crystalline states.* SAXS was used to probe the nanostructure of PE vitrimers because of the large scattering contrast between the backbone and graft. For an X-ray wavelength of 1 Å, PE and dioxaborolane maleimide have a calculated scattering contrast of $1 \times 10^{-7}$ nm$^{-4}$ – 6 times higher than the calculated scattering contrast between polystyrene and poly(methyl methacrylate).[81] At 160 °C (~30 °C above the PE melting temperature), both neat and peroxide crosslinked PE melts display SAXS scattering patterns characteristic of a homogeneous material (Figure S11). In striking contrast, molten PE-*g*, PE-*v*, and insoluble portion of PE-*v* SAXS patterns all feature a 2-shoulder scattering signature (Figures 8A, 8B, and S12). The low- and high-$q$ shoulders begin to descend at ~0.02 and ~0.2 nm$^{-1}$, respectively. The high-$q$ scalings in the SAXS patterns (*i.e.*, the slope of $I$ vs. wavevector $q$ in logarithmic axes) vary between -3 and -4, lower than the expected -2 scaling for scattering from a gel.[82,83] These scalings, also inconsistent with scattering models for a statistical copolymer correlation hole,[84,85,86] suggest that the PE-*g* and PE-*v* nanostructures display sharp interfaces. Remarkably, the initial and insoluble portion of PE-*v* patterns are almost identical; the insoluble materials just have higher intensity (Figures 8C and S13). If the initial PE-*v* pattern is vertically shifted, it completely overlaps the insoluble portion of PE-*v* pattern (Figures 8D and S14), indicating that the length scales of the PE-*v* nanostructure are undisturbed by the removal of the soluble portion. Long-time annealing experiments and annealing at different temperatures well above the melting point show that the observed melt SAXS patterns can be stable for at least 48 hrs and are insensitive to temperature (Figures S15).

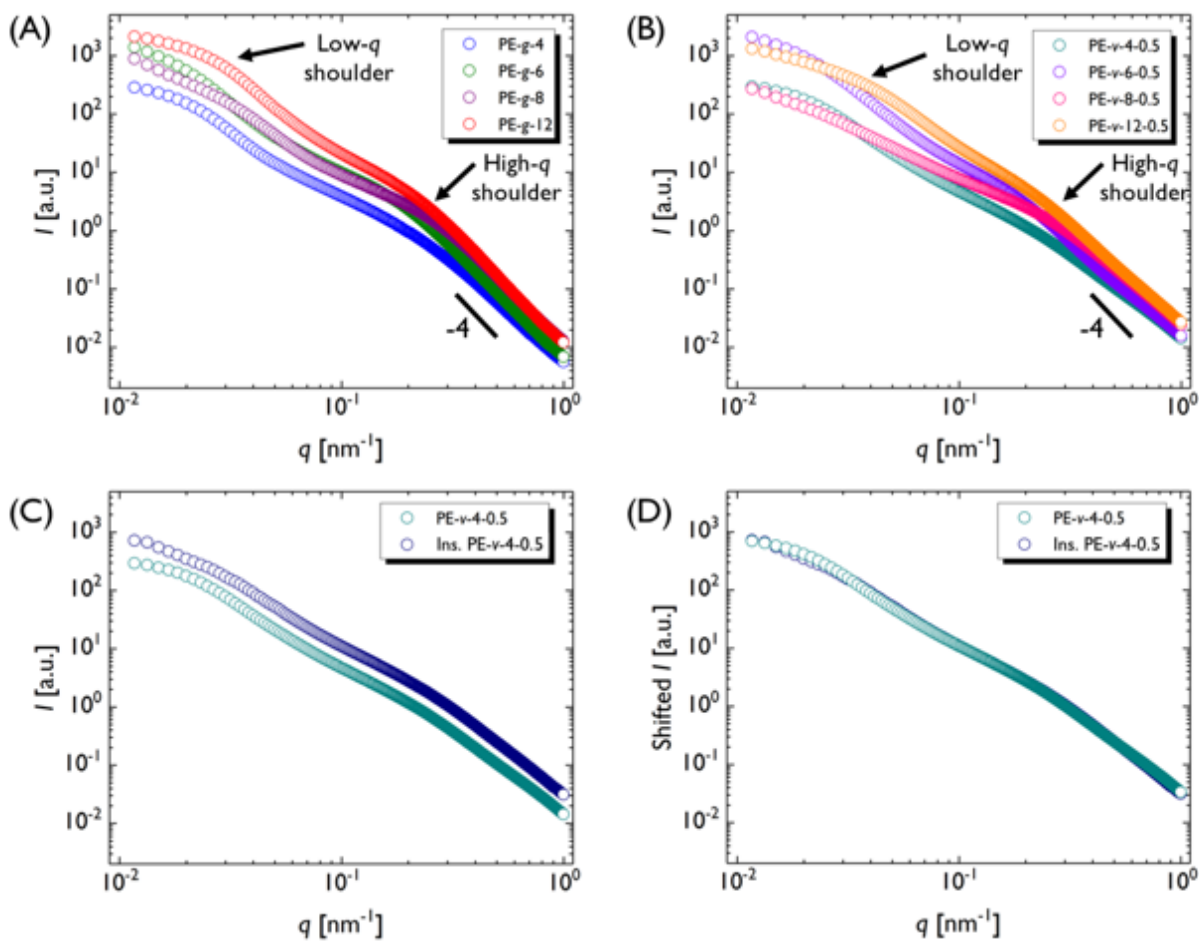

**Figure 8.** Small-angle X-ray scattering patterns (SAXS) of (A) PE-*g* and (B) PE-*v* samples at 160 °C. (C) Comparison of SAXS patterns for initial and insoluble portion of PE-*v*-4-0.5 at 160 °C. (D) The initial PE-*v*-4-0.5 SAXS pattern is vertically shifted onto the insoluble portion of PE-*v*-4-0.5 SAXS pattern.

Semi-crystalline PE-*g* and PE-*v* (*i.e.*, materials cooled from 160 to 40 °C at a rate of -10 °C/min) also have more complex nanostructures than the neat and peroxide crosslinked PE. SAXS patterns revealed that along with the crystal lamellae structure factor, semi-crystalline PE-*g* and PE-*v* exhibited the same low-*q* shoulder observed in the melt SAXS patterns (Figure 9A). The high-*q* shoulder is likely hidden by the crystal lamellae structure factor. If the melt SAXS pattern of a particular PE-*g* or PE-*v* sample is vertically shifted onto the corresponding semi-crystalline SAXS pattern, the low-*q* shoulder regions completely overlap (Figures 9B and S16-S18), implying that the low-*q* length scales in the melt and semi-crystalline states are identical. Wide-angle X-ray scattering and electron density correlation function analysis show that the presence of the grafts and crosslinks do not alter the PE crystallite unit cell parameters and average lamella thickness (Figures S19-S22 and Table S2). Polarized optical microscopy, however, reveals that PE-*g* and PE-*v* do not form spherulites (Figure S23).

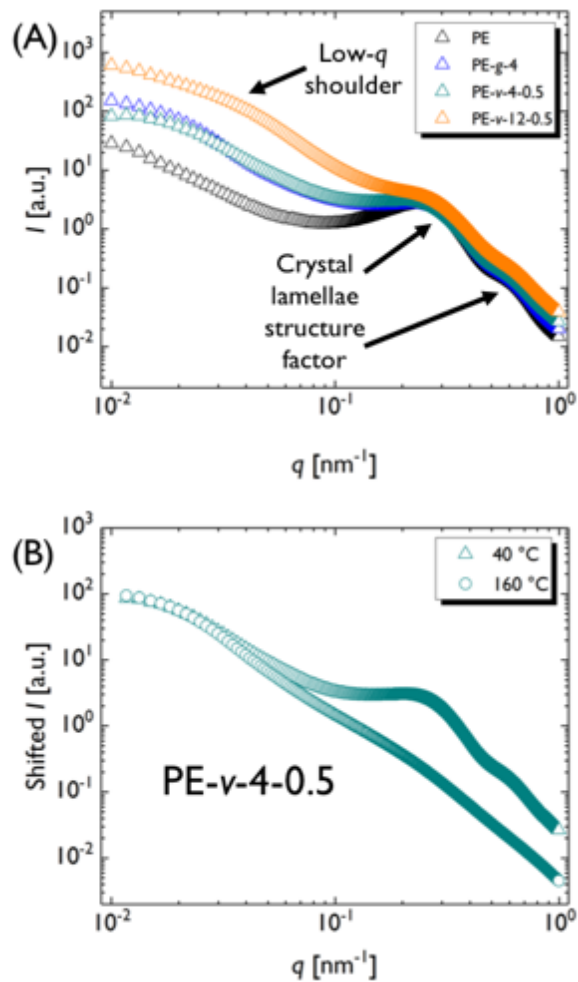

**Figure 9**. (A) Comparison of SAXS patterns for various semi-crystalline PE, PE-*g*, and PE-*v* samples at 40 °C. (B) Comparison of PE-*v*-4-0.5 SAXS patterns at 40 and 160 °C. The 160 °C pattern was vertically shifted to show the overlap in the low-*q* shoulder region.

The analyses of the molten and semi-crystalline SAXS patterns demonstrate that vitrimers undergo complex nanoscale self-assembly. Inspired by studies of PE ionomers and other associating polymer systems, the SAXS patterns were hypothesized to be evidence of hierarchical micro-phase separation between the PE backbone and grafts. According to this hypothesis, the high-$q$ shoulder represents scattering from graft-rich aggregates. Clustering of the grafts likely occurs to minimize the enthalpic penalty of PE/dioxaborolane maleimide dispersion interactions. The aggregate may be described as a sphere with radius $R_{sph}$, and the lack of oscillations in the patterns hints that the aggregates are polydisperse. PE strands fill the space between aggregates; they can form loops or connect different clusters. Regardless of the chain conformations, SAXS captures the amorphous PE phase as a matrix for the aggregates. The lack of a sharp Bragg peak in the SAXS patterns conveys that the aggregates are not densely packed, which contrasts the observations of PE ionomer systems.[56] Instead, the low-$q$ shoulder suggests that the aggregates assemble into an ill-defined fractal shape with fractal dimension $D$ and length scale $\xi$.

To test the hypothesized hierarchical nanostructure, the PE-$g$, PE-$v$, and insoluble portion of PE-$v$ SAXS patterns were fit to an aggregate-fractal scattering model. The model features $P_{sph}$, the form factor of polydisperse spheres (Equations S8–S11), and $S_{frac}$, a mass fractal structure factor with functional form

$$S_{frac} = 1 + \alpha'_{frac} \frac{\sin[(D-1)\tan^{-1}(q\xi)]}{(D-1)q\xi(1+q^2\xi^2)^{(D-1)/2}} \qquad (5)$$

where $q$ is the scattering vector, $D$ is the fractal dimension, $\xi$ is the fractal length scale, and $\alpha'_{frac}$ is an amplitude factor that describes the scattering contrast between the fractal components and the matrix.[87,88] Because the low- and high-$q$ shoulder onsets differ by at least an order of magnitude, the total scattering intensity was approximated to be the sum of scattering from the mass fractals and the individual aggregates:

$$I(q) = \Delta\rho\phi P_{sph}S_{frac} \approx \alpha_{sph}P_{sph}(q; R_{sph}, \sigma) + \alpha_{frac}\frac{\sin[(D-1)\tan^{-1}(q\xi)]}{(D-1)q\xi(1+q^2\xi^2)^{(D-1)/2}} \quad (6)$$

where $R_{sph}$ is the average aggregate radius, $\sigma$ is the aggregate radius standard deviation, and $\alpha_{sph}$ and $\alpha_{frac}$ are the amplitude factors for scattering from an individual aggregate and mass fractal, respectively.

To minimize the number of free-floating parameters during the model fitting, two assumptions were made. First, the aggregate size was assumed to be less than or equal to the amorphous layer thickness of the semi-crystalline materials. This assumption relies on the observations that the nanostructure length scales were the same in both the melt and semi-crystalline states, and that the grafts and crosslinks persist in the amorphous fraction. Presumably, formation of crystal lamellae pushes the aggregates into the amorphous layers (see Figure S24). To be consistent with this physical picture, the upper bound for $R_{sph}$ should be half the measured amorphous layer thickness (*i.e.*, 1.5–2.5 nm). The aggregate-fractal model also does not fit the data if $R_{sph}$ < 1.5 nm. Therefore, $R_{sph}$ was fixed to be 2 nm for all samples. Second, $\alpha_{sph}$ was assumed to be linearly proportional to the graft density. The justification for this assumption stems from preliminary model fits to a 2-term Debye-Bueche scattering model (Figure S25 and Table S3), which revealed that the high-*q* shoulder scattering contribution was linearly proportional to grafting density. Increasing graft density likely increases the number of aggregate scatterers. Invoking this assumption, preliminary free fits to the aggregate-scattering model were used to develop a calibration curve that enabled $\alpha_{sph}$ to be calculated *a priori* (Figure S26). By limiting both $R_{sph}$ and $\alpha_{sph}$, the model fits were performed using only four free-floating parameters: $\sigma$, $\alpha_{frac}$, $D$, and $\xi$.

The aggregate-fractal scattering model describes the PE-*g*, PE-*v*, and insoluble portion of PE-*v* SAXS data very well (Figures 10A and S27). Figure S28A shows that the fitted values for $\sigma$ are essentially the same for all samples (~1.4 nm), but the estimated values for the fractal parameters hint at interesting structure relationships. $\alpha_{frac}$ and $D$ seem to be sensitive to the reactive mixing and extrusion feed composition (Figures S28B and S28C). PE-*g*-8 and PE-*v*-8-0.5 have similar $\alpha_{frac}$ and $D$ values as the feed PE-*g*-4 that was used for their synthesis. The same relationship exists among PE-*g*-12, PE-*v*-12-0.5, and the feed PE-*g*-6. This phenomenon may possibly be related to the distribution of the dioxaborolane maleimide grafts along the feed PE-*g* backbone. Lowering the ratio between crosslinker and graft also alters $\alpha_{frac}$ (Figure S29). As seen in Figure 10B, $\xi$ is similar between a particular PE-*g* sample and its corresponding PE-*v* specimens (both initial and insoluble material). This length scale, which decreases as the precursor PE-*g* grafting density increases, ranges from 70–30 nm. Because the measured values are much larger than the radius of gyration for a single PE chain (~10 nm), $\xi$ does not describe the distance between two aggregates. Instead, $\xi$ is commensurate with the PE crystal lamellae dimensions. As revealed by the electron density correlation functions, the average lamella thickness is about 16 nm, while the lateral dimensions are likely larger. The resemblance between these length scales suggests that $\xi$ is related to the crystallization of the PE backbone.

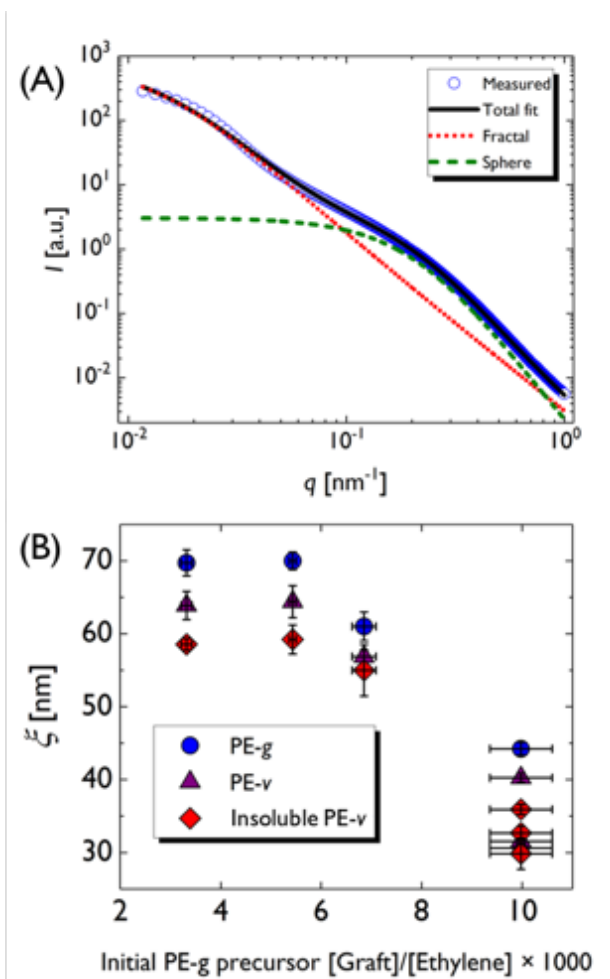

**Figure 10**. (A) SAXS pattern of PE-*g*-4 fit to the aggregate-fractal scattering model. (B) Fractal length scale $\xi$ for PE-*g*, PE-*v*, and insoluble portion of PE-*v* samples.

Based on X-ray scattering analyses, unfavorable enthalpic interactions between PE and the dioxaborolane maleimide graft seem to drive PE-*g* and PE-*v* to self-assemble into hierarchical nanostructures. The discovery of these structures, which persist in both the molten and semi-crystalline states, illustrates the rich polymer physics of vitrimer materials made of interacting strands and dynamic crosslinks.

**Discussion and conclusions**

Figure 11 presents a schematic of the proposed PE vitrimer morphology. At micron length scales, macroscopic phase separation into domains rich and poor in grafts/crosslinkers occurs. Consequently, all vitrimer samples investigated in this work were turbid at temperatures above the melting point. At the nanoscale, the grafts/crosslinks form small aggregates, as reflected by the high-$q$ shoulder in the SAXS patterns. At low-$q$ values, the SAXS patterns reveal that the aggregates assemble to form a fractal object $\geq$ 30 nm in size and with a fractal dimension between 2.5–3.0. Because grafted and/or dynamically crosslinked PE chains may possess several dioxaborolane maleimide units, a given chain can form a loop within a single aggregate or serve as a bridge between two aggregates.

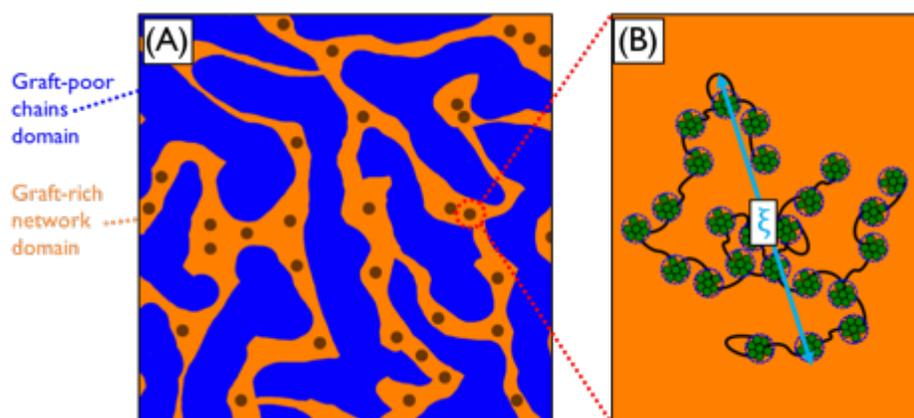

**Figure 11.** Schematic of proposed (A) meso- and (B) nanostructure of molten PE-*g* and PE-*v* materials. In (A), dark circles represent fractals. In (B), aggregates, PE chain bridges, and PE chain loops are shown. $\xi$ is the fractal length scale.

Macro- and micro-phase separation also occur in the semi-crystalline state. The DSC results suggest that the overall PE crystallinity differs between the crosslinked graft-rich and graft-poor phases. Remarkably, however, the crystallite dimensions (*i.e.*, unit cell parameters, periodicity, and lamellae and amorphous layer thicknesses) in neat PE, PE-*g*, and PE-*v* are basically the same. This hints that crystallization of PE located between aggregates is more difficult, but crystallites that do form are just like the ones in regular PE. The lack of an effect likely stems from the relatively sparse graft densities of the samples. In molar terms, the grafting densities explored in this study are relatively low (< 2.5 mol%). Moreover, both aggregation and formation of fractal assemblies of aggregates leave large regions with no dioxaborolane maleimide. In these regions, PE can readily crystallize. The relatively little molar amounts of grafting in these samples, however, are enough to inhibit spherulite growth. The complete elimination of spherulites is also seen in PE ionomer systems, where clustering of the ions is thought to reduce the lateral lamellae dimensions and disturb the lathlike growth of spherulites.[89,90] A similar mechanism potentially occurs in the PE-$g$ and PE-$v$ materials. The similarity between the fractal length scale $\xi$ and the PE crystallite dimensions suggests that crystallization also influences the aggregate packing. Overlap of the low-$q$ region in the molten and semi-crystalline SAXS patterns shows that the $\xi$ values of the two states are identical, hinting that the aggregate packing arrangement must be able to accommodate the growth of crystallites.

The time scale for relaxation of the aggregates into a more homogeneous packing is likely exceptionally long (>> 48 hrs). Many kinetic obstacles operate even in the molten state – *e.g.*, enthalpic barrier of transporting a dioxaborolane maleimide moiety through a PE-rich region, exchange reactions occurring mostly within a single aggregate, and PE chain entanglements. Thus, memory effects of sample synthesis and preparation are expected to be difficult to erase by simple thermal annealing.

Additional studies are needed to elucidate the details of the nanostructure dimensions and dynamics, but the implication of the current findings is clear: self-assembly must be considered when designing vitrimer systems. Similar to block polymers and ionomers, micro-phase separation in vitrimers can be potentially exploited to enhance mechanical and rheological properties. Specifically, the melt dynamics could be optimized for a particular application or process by balancing the exchange reaction kinetics and crosslink association strength. Macro-phase separation may also be employed to aid in vitrimer processing. For instance, the astonishingly short stress relaxation times of PE vitrimers potentially result from the graft-poor macrophase acting as a lubricant for the graft-rich percolating network.[23]

**Author information**

*Corresponding Authors*

*E-mail: ralm.ricarte@espci.fr and ludwik.leibler@espci.fr

*ORCID*

Ralm Gerick Ricarte: 0000-0003-1018-6083

*Notes*

The authors have no competing financial interest.

**Acknowledgements**

We are very grateful to Pierre-Antoine Albouy and Doru Constantin for helping us perform preliminary lab-scale SAXS measurements. Synchrotron-sourced small-angle X-ray scattering experiments were performed on beamline ID-02 at the European Synchrotron Radiation Facility (ESRF), Grenoble, France. We are grateful to Lewis Sharpnack at the ESRF for providing assistance in using beamline ID-02. This research used resources of the Advanced Photon Source, a U.S. Department of Energy (DOE) Office of Science User Facility operated for the DOE Office of Science by Argonne National Laboratory under Contract No. DE-AC02-06CH11357. We thank Denis T. Keane of the Advanced Photon Source for his help in using beamline DND-CAT 5ID-D. We acknowledge SOLEIL for provision of synchrotron radiation facilities and we would like to thank Thomas Bizien for assistance in using beamline SWING. We would also like to acknowledge helpful discussions with Trystan Domenech, Rob van der Weegen, and Julie Kornfield. We are indebted to Esther Cazares-Cortes, Sarah Goujard, Mickaël Pomes-Hadda, Takeshi Kondo, Maïka Saint-Jean, and Jeffrey Ting for their assistance in conducting the SAXS measurements. The research leading to these results has received funding from the People Programme (Marie Curie Actions) of the European Union's Seventh Framework Programme (FP7/2007-2013) under REA grant agreement n. PCOFUND-GA-2013-609102, through the PRESTIGE programme (2017-2-0012) coordinated by Campus France.**Supporting information**

Preparation of peroxide crosslinked polyethylene, determination of absorption coefficients for composition quantification, optical transmission instrument schematic and turbidity calculation, comparison of FTIR spectra for PE-*v* materials with varying crosslink density, decomposed FTIR spectra and estimated graft and crosslink densities for PE-*g* and PE-*v*, estimated graft distributions for PE-*g* and PE-*v*, phase contrast optical microscopy, turbidity and phase contrast optical microscopy of insoluble portion of PE-*v* samples, differential scanning calorimetry data of PE, peroxide crosslinked PE, PE-*g*, PE-*v*, and insoluble portion of PE-*v* samples, small-angle X-ray scattering pattern background removal, comparison of initial PE-*v* and insoluble portion of PE-*v* SAXS patterns, influence of annealing time and temperature on SAXS patterns, comparison of SAXS patterns at 40 and 160 °C, wide-angle X-ray scattering patterns, PE crystallite dimensions quantification via electron density correlation function analysis, polarized optical microscopy, hypothesized semi-crystalline nanostructure for PE-*g* and PE-*v*, SAXS pattern fits to the 2 term Debye-Bueche scattering model, and SAXS pattern fits to the aggregate-fractal scattering model are provided in the Supporting Information.

**TOC Image**

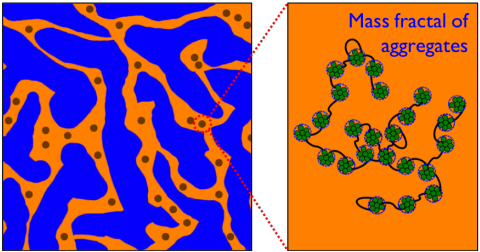

Supporting information for

**Phase separation and self-assembly in vitrimers: hierarchical morphology of molten and semi-crystalline polyethylene/dioxaborolane maleimide systems**

Ralm G. Ricarte[*,†], François Tournilhac[†], and Ludwik Leibler[*,‡]

[†]Matière Molle et Chimie, École Supérieure de Physique et de Chimie Industrielles de la Ville de Paris (ESPCI)–CNRS, UMR-7167, Paris Sciences et Lettres (PSL) Research University, 10 Rue Vauquelin, 75005 Paris, France and [‡] UMR CNRS 7083 Gulliver, ESPCI Paris, PSL Research University, 10 Rue Vauquelin, 75005 Paris, France

Email: ralm.ricarte@espci.fr and ludwik.leibler@espci.fr

*Table of contents*

*Preparation of peroxide crosslinked polyethylene*

Peroxide crosslinked polyethylene was prepared by reactive mixing followed by extrusion. The procedure was performed using a DSM Explore batch twin-screw extruder (capacity of 5 cm$^3$) with a recirculation channel and a co-rotating conical screw profile. First, neat polyethylene (PE) and 3 wt% of dicumyl peroxide were mixed in polypropylene bottles. The mixture was charged into the extruder, which was set to a temperature of 170 °C and a screw speed of 100 rpm. After a residence time of less than 10 s, the mixture was extruded, directly placed into a disc-shaped mold (diameter: 25 mm; thickness: 1.5 mm), and immediately compression molded at 150 °C. Transfer time between the extruder and compression molder was less than 60 s. The resulting peroxide crosslinked PE had an insoluble fraction of 53 wt% of PE.

*Determination of absorption coefficients for composition quantification*

Solution Fourier transform infrared spectroscopy (FTIR) measurements were performed to determine the absorption coefficients ($\varepsilon$) of characteristic signals required for composition quantification of graft-functionalized PE (PE-*g*) and PE vitrimer (PE-*v*). The measurements were made using a Bruker Tensor 37 spectrometer equipped with a Specac® Omni-cell Liquid Transmission Cell with KBr windows. Solutions featuring chloroform as the solvent and different concentrations of either dioxaborolane maleimide, bis-dioxaborolane crosslinker, or docosane (an oligomer analogue of PE) were analyzed. The observed peaks in the spectra of these small molecule solutions matched the characteristic signals seen in the spectra of the PE, PE-*g*, and PE-*v* materials (Figure S1). For each spectrum, the areas under the peaks of interest were measured and plotted as a function of solute concentration (Figure S2). Because each peak exhibited a linear relationship with the solute molar concentration, the $\varepsilon$ values of the characteristic signals were calculated using the Beer-Lambert law.

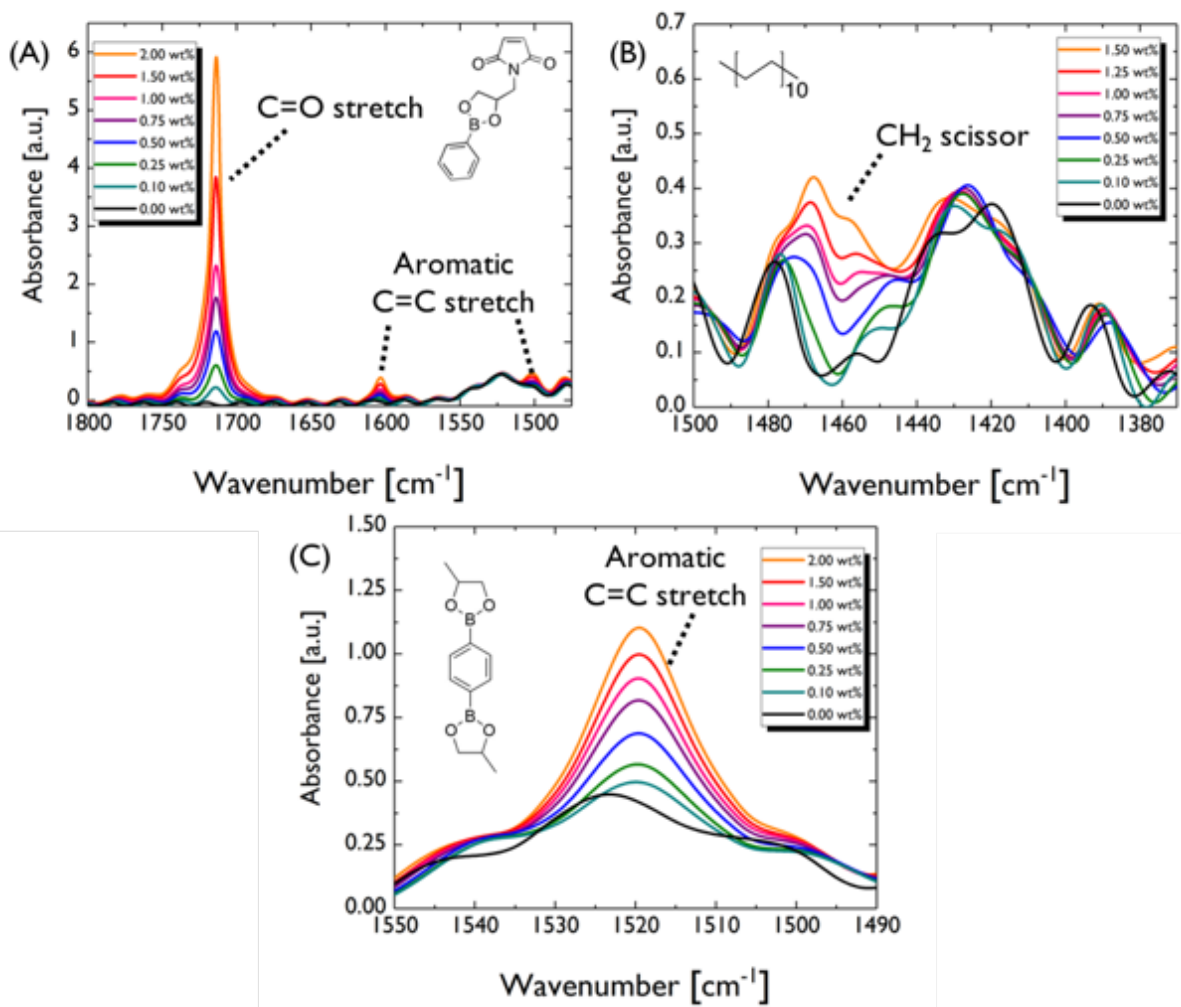

**Figure S1.** FTIR spectra for solutions of (A) dioxaborolane maleimide, (B) docosane, and (C) bis-dioxaborolane dissolved in chloroform.

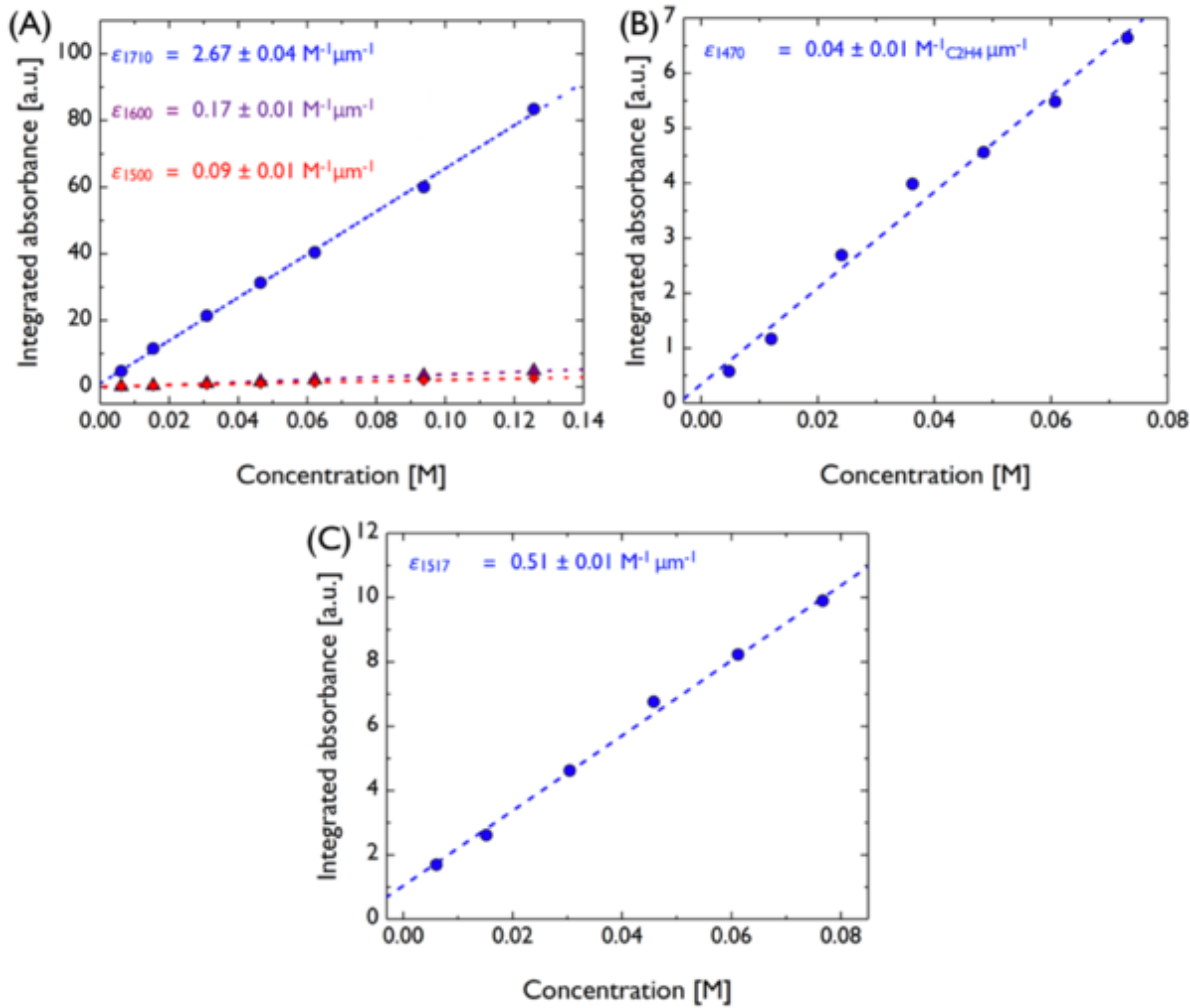

**Figure S2.** Measured absorption coefficients ($\varepsilon$) for characteristic FTIR signals of (A) dioxaborolane maleimide, (B) docosane, and (C) bis-dioxaborolane crosslinker. $\varepsilon_{1470}$ is reported on a per mole of $C_2H_4$ basis. All other $\varepsilon$ values are reported on a per mole of the entire molecule basis. Error bars of the reported values are the standard error of the linear regression.

*Optical transmission instrument schematic and turbidity calculation*

Figure S3 is the schematic for the custom-built optical transmission instrument used for measuring the turbidity of the PE-*g* and PE-*v* materials.

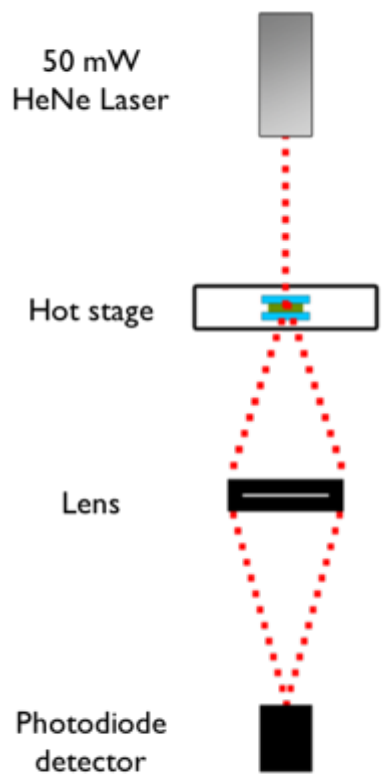

**Figure S3.** Schematic of custom-built optical transmission instrument.

Equation S1 was used to calculate the turbidity, $\delta$, of the molten PE-*g* and PE-*v* materials:

$$\delta = \frac{\ln(I_0/I_t)}{L} \qquad (S1)$$

where $I_0$ is the incident intensity, $I_t$ is the transmitted intensity, and $L$ is the sample thickness.[S1]

*Comparison of FTIR spectra for PE-v materials with varying crosslink density*

Figure S4 compares the FTIR spectra of PE-*v*-12-0.5, PE-*v*-12-0.3, and PE-*v*-12-0.2. These samples have approximately the same graft density, but varying crosslink densities. The spectrum for PE-*v*-12-0.5 exhibits the aromatic C=C stretch signal associated with the bis-dioxaborolane crosslinker, but does not display the aromatic C=C stretch peaks characteristic to the phenyl ring of the dioxaborolane maleimide graft. The absence of these peaks suggests that the grafts are fully saturated by crosslinker. However, it does not distinguish between the presence of crosslink that acts as a junction point or a dangling bis-dioxaborolane. Conversely, the spectra for PE-*v*-12-0.3 and PE-*v*-12-0.2 show all three aromatic C=C signals, insinuating that these materials contain grafts unsaturated by the crosslinker. The quantified crosslink densities for these materials are listed in Table 1 in the main text.

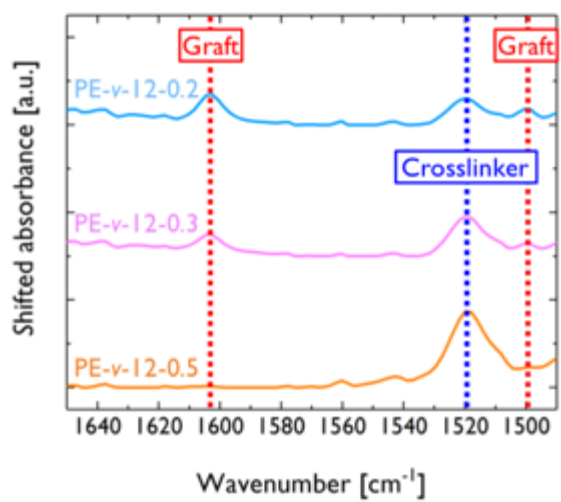

**Figure S4.** Comparison of aromatic C=C signals in FTIR spectra for PE-*v*-12-0.5, PE-*v*-12-0.3, and PE-*v*-12-0.2. Spectra were shifted for clarity.

*Decomposed FTIR spectra and estimated graft and crosslink densities for PE-g and PE-v*

Figure S5 displays the decomposed FTIR spectra for the PE-*g* and PE-*v* samples. Each characteristic peak was fit to a skewed Gaussian equation.

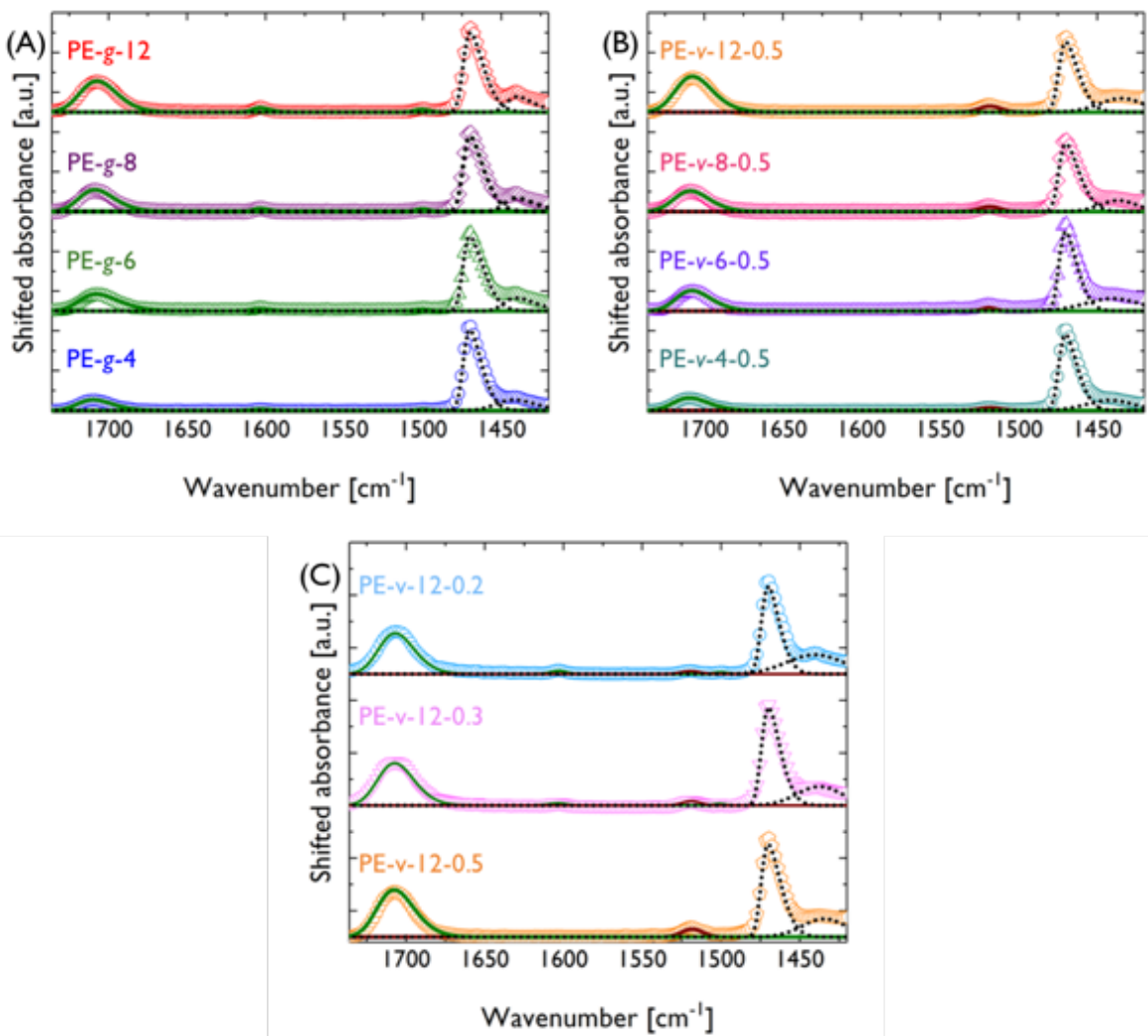

**Figure S5.** Decomposed FTIR spectra for samples of (A) PE-*g*, (B) PE-*v* with varying graft density, and (C) PE-*v* with varying crosslink density. The PE-*v*-12-0.5 spectra in (B) and (C) are the same. The black, green, and brown fit lines correspond to signals from the PE backbone, dioxaborolane maleimide graft, and bis-dioxaborolane crosslinker, respectively. Spectra were shifted for clarity.

*Estimated graft distributions for PE-g and PE-v*

The large molecular weight dispersity of the initial PE ($Đ$ = 7.5, see Figure 2) causes the number of grafts per chain distribution to be broad. To estimate this distribution for each sample, a simple statistical model – which assumes dioxaborolane maleimide adds to the PE backbone in a random manner – was employed.[S2] In this model, $N$ is the degree of polymerization of the chain and $N_s$ is the number of grafts on the chain backbone. The number fraction of chains with $N$ segments and $N_s$ grafts, $\phi$, may be described as the product of two separate distributions

$$\phi(N, N_s) = \theta(N) \times \psi(N, N_s) = \theta(N) \frac{N!}{(N - N_s)! N_s!} p^{N_s} (1 - p)^{N - N_s} \qquad (S2)$$

where $\theta$ is the molecular weight distribution of all PE chains, and $\psi$ is the $N_s$ distribution for chains with length $N$. $\psi$ was estimated using the binomial distribution, where $p$ is the average mole fraction of grafts. Because all chains encompassed by $\theta(N)$ have identical degrees of polymerization, the number and weight fractions of chains with different $N_s$ are equivalent. $\theta$ was determined using the high-temperature size-exclusion chromatography trace of the neat high-density PE (see Figure 2). The calculated distributions from this model indicate that the PE-*g* samples not only have a wide graft number distribution, but also that they contain a considerable fraction of chains that have no grafts or well below the global average (Figures S6A–S6D). Figure S6E details the weight fraction of chains with 0 grafts for each sample. For PE-*v*, the non-functionalized chains are not incorporated into the network; they act as a diluent and can be a source of macroscopic phase separation.

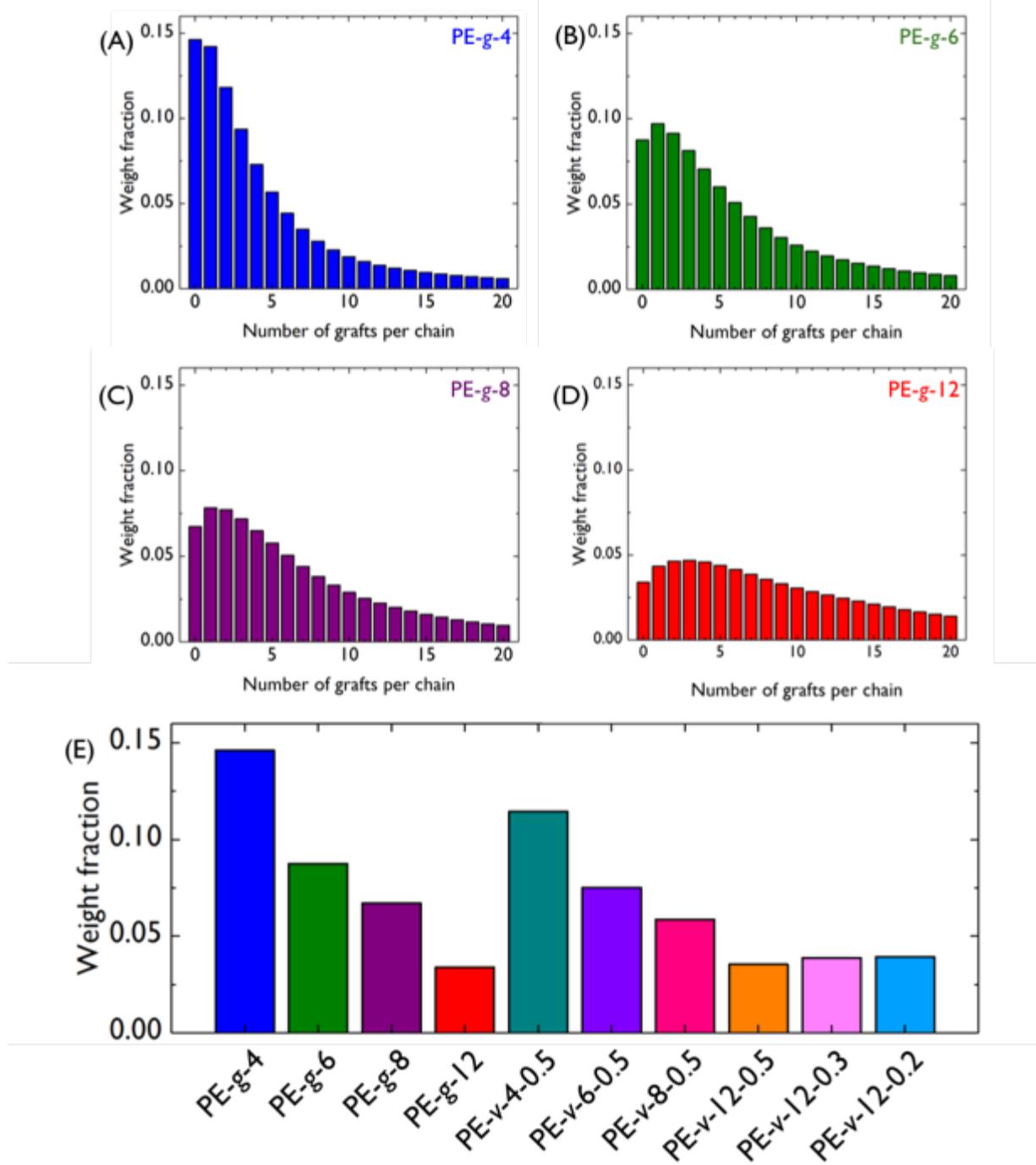

**Figure S6.** Estimated number of grafts per chain distribution for (A) PE-*g*-4, (B) PE-*g*-6, (C) PE-*g*-8, and (D) PE-*g*-12. (E) Estimated weight fraction of chains with 0 grafts for PE-*g* and PE-*v* samples.

*Phase contrast optical microscopy*

Phase contrast optical microscopy was performed using a LEICA DMR DAS microscope equipped with a 10×/0.30 HC PL Fluotar objective lens. A 150 μm thick film of the sample was put between a glass microscope slide and glass coverslip, and then placed into a Linkam Scientific LTS350 heating stage. The sample was annealed at 160 °C for at least 5 minutes before phase contrast micrographs were recorded. Figure S7A shows that neat molten PE, a homogeneous material, has a uniform surface contrast. In contrast, peroxide crosslinked PE has a more heterogeneous surface contrast that is attributed to surface roughness (Figure S7B). PE-*g*-4 also displays heterogeneous surface contrast (Figures S7C). When dynamic crosslinker is added to the system, the surface morphology becomes strikingly different. As seen in Figure S7D, PE-*v* materials exhibit a rugged surface morphology with well-defined interfaces. This observation suggests that the combination of macro-phase separation and dynamic crosslinking causes vitrimers to undergo a syneresis-like phenomenon. The brighter regions are interpreted to be a graft-rich percolating network, while the darker regions are viewed as deficient in graft.

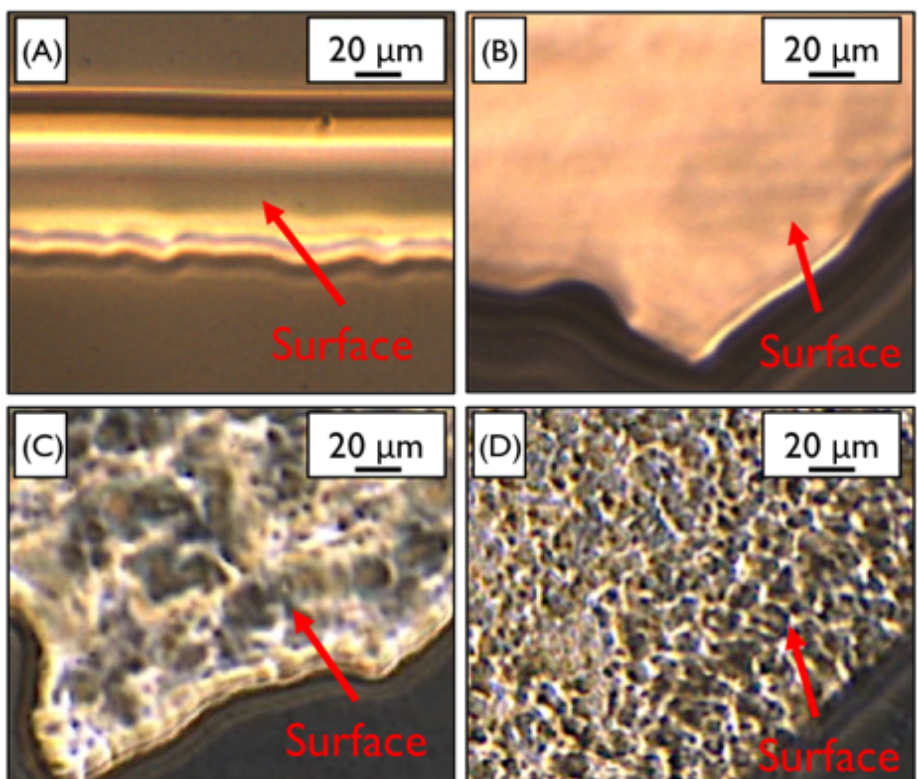

**Figure S7.** Phase-contrast optical microscopy images of (A) neat PE, (B) peroxide crosslinked PE, (C) PE-*g*-4, and (D) PE-*v*-4-0.5. In all images, the region-of-interest is an area where the optical contact is poor (and thus, image contrast is high). In (A), the red arrow is pointing to the edge of the sample. In (B), (C), and (D) the red arrows are pointing to regions of the film that have delaminated from the glass slide.

S11

*Turbidity and phase contrast optical microscopy of insoluble portion of PE-v samples*

Figure S8A features the measured turbidities of PE, peroxide crosslinked PE, and insoluble portion of PE-*v*. The slight increase of the peroxide crosslinked PE turbidity relative to the neat polymer is attributed to surface roughness. Unlike the PE-*g* and PE-*v* samples, the turbidity of the insoluble portion of PE-*v* samples is not linearly proportional to the graft density. Figure S8B is a phase contrast optical microscopy image of the insoluble portion of PE-*v*-4-0.5. Although the surface of the material still retains a rugged morphology, the feature sizes are smaller than those seen in the initial vitrimer.

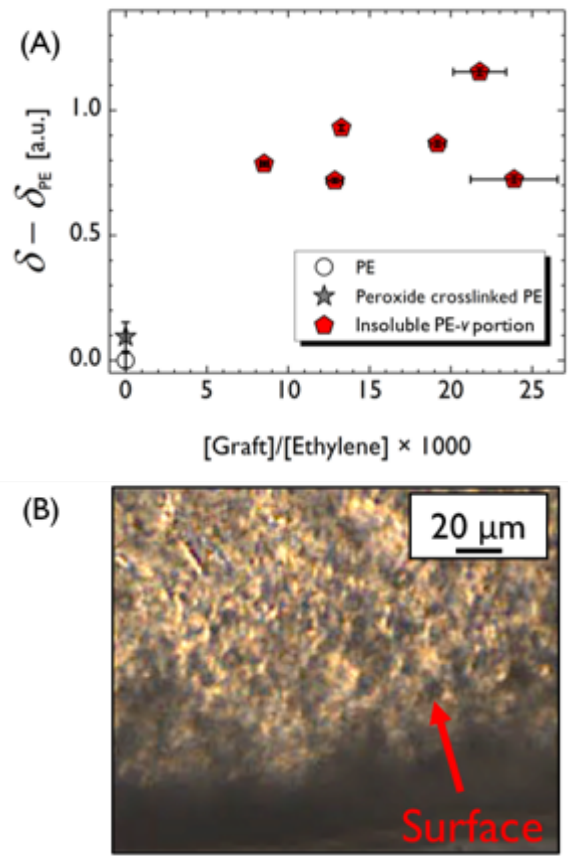

**Figure S8.** (A) Turbidity measurements for PE, peroxide crosslinked PE, and insoluble portion of PE-*v*. All samples were 1 mm thick. Y-axis error bars are the propagated error. (B) Phase-contrast optical microscopy image of insoluble portion of PE-*v*-4-0.5. The arrow refers to an area that has delaminated from the glass side and exhibits enhanced light scattering contrast.

*Differential scanning calorimetry data of PE, peroxide crosslinked PE, PE-g, PE-v, and insoluble portion of PE-v samples*

Figure S9 features the first cooling and second heating DSC traces for PE, peroxide crosslinked PE, PE-*g*, PE-*v*, and insoluble portion of PE-*v* samples. Figure S10 exhibits the overall crystallinity values, while Table S1 tabulates the $T_c$, $T_m$, and $\Delta H_m$ for all samples.

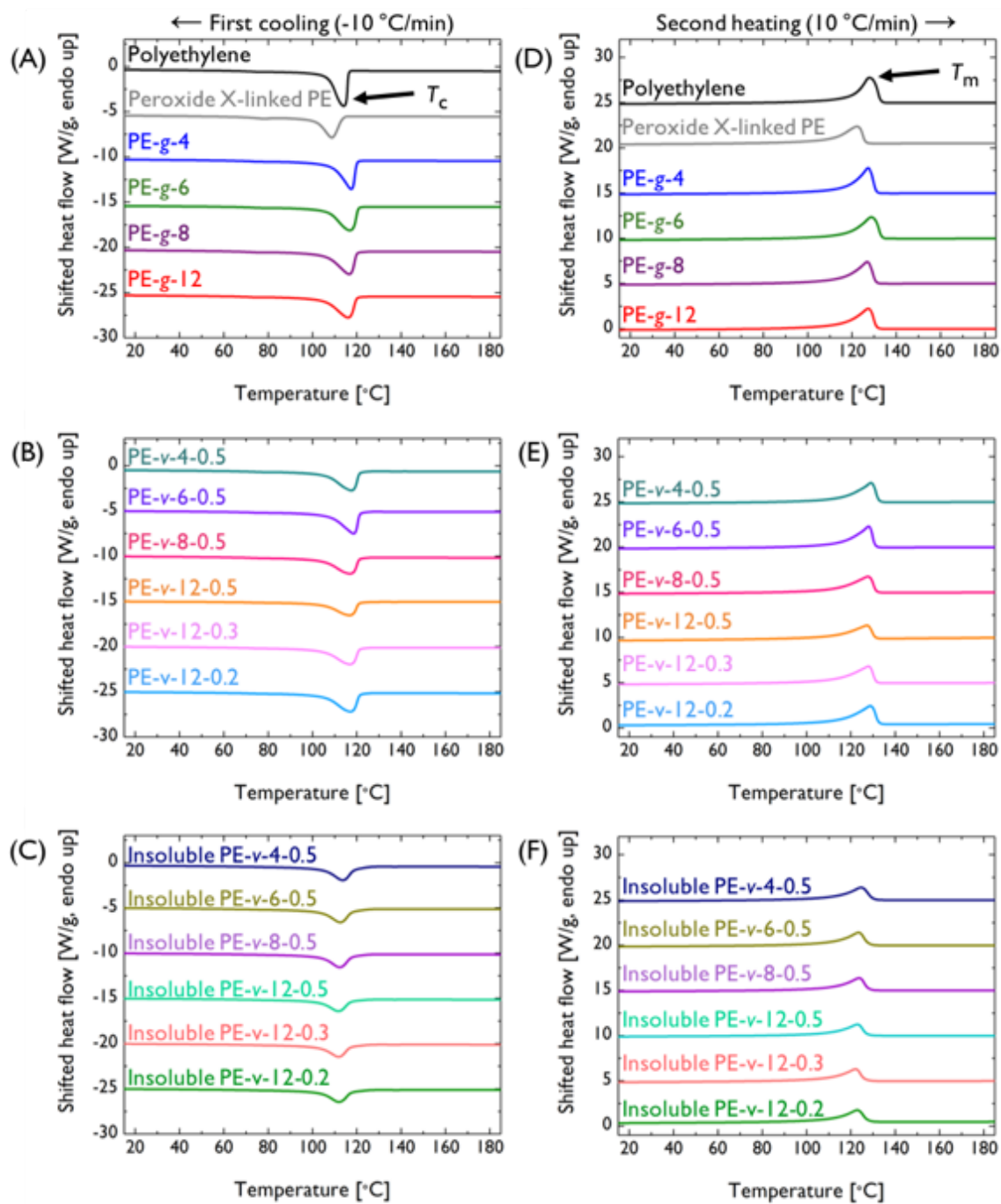

**Figure S9.** DSC traces for PE, peroxide crosslinked PE, PE-*g*, PE-*v*, and insoluble portion of PE-*v*. Panels (A)–(C) are the first cooling traces, and (D)–(F) are the second heating traces. Traces were shifted for clarity.

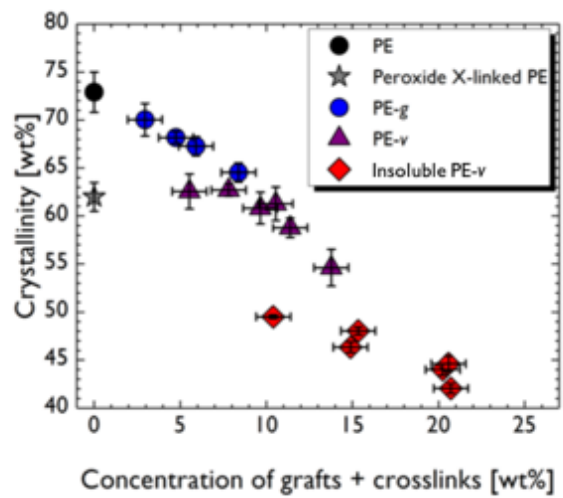

**Figure S10.** Overall crystallinity for PE, peroxide crosslinked PE, PE-*g*, PE-*v*, and insoluble portion of PE-*v*. Y-axis error bars represent the standard deviation of three different measurements.

**Table S1.** Crystallization temperature ($T_c$), melting temperature ($T_m$), and melting enthalpy ($\Delta H_m$) values. Error bars are the standard deviation of three different measurements.

| Sample | $T_c$ [°C] | $T_m$ [°C] | $\Delta H_m$ [J/g] |
|---|---|---|---|
| PE | 114 ± 1 | 128 ± 1 | 221 ± 6 |
| Peroxide crosslinked PE | 109 ± 1 | 120 ± 1 | 187 ± 5 |
| PE-g-4 | 116 ± 1 | 128 ± 1 | 212 ± 5 |
| PE-g-6 | 117 ± 1 | 129 ± 1 | 207 ± 2 |
| PE-g-8 | 116 ± 1 | 127 ± 1 | 204 ± 1 |
| PE-g-12 | 116 ± 1 | 127 ± 1 | 196 ± 2 |
| PE-v-4-0.5 | 118 ± 1 | 128 ± 1 | 190 ± 5 |
| PE-v-6-0.5 | 118 ± 1 | 128 ± 1 | 190 ± 2 |
| PE-v-8-0.5 | 117 ± 1 | 128 ± 1 | 184 ± 5 |
| PE-v-12-0.5 | 116 ± 1 | 128 ± 1 | 166 ± 6 |
| PE-v-12-0.3 | 118 ± 1 | 128 ± 1 | 178 ± 3 |
| PE-v-12-0.2 | 118 ± 1 | 128 ± 1 | 186 ± 5 |
| Insoluble portion of PE-v-4-0.5 | 114 ± 1 | 124 ± 1 | 150 ± 1 |
| Insoluble portion of PE-v-6-0.5 | 113 ± 1 | 124 ± 1 | 146 ± 2 |
| Insoluble portion of PE-v-8-0.5 | 112 ± 1 | 124 ± 1 | 140 ± 2 |
| Insoluble portion of PE-v-12-0.5 | 112 ± 1 | 123 ± 1 | 127 ± 2 |
| Insoluble portion of PE-v-12-0.3 | 112 ± 1 | 122 ± 1 | 135 ± 2 |
| Insoluble portion of PE-v-12-0.2 | 112 ± 1 | 123 ± 1 | 133 ± 2 |

*Small-angle X-ray scattering pattern background removal*

All small-angle X-ray scattering (SAXS) patterns presented in this manuscript were first processed by subtracting a background pattern of an empty sample cell sealed with two layers of 25 μm Kapton film. The SAXS patterns of molten neat and peroxide crosslinked PE after this first background removal are presented in Figure S11A. Although both these materials are homogeneous, they still display a continuous background of liquid-like scattering that follows a scattering vector ($q$) power law scaling of -3.15. This type of pattern, seen for all homogeneous materials, is typically attributed to electron density fluctuations, parasitic scattering from the beamline equipment, or unknown homogeneities within the sample.[S3,S4] The SAXS patterns of the molten PE-*g*, PE-*v*, and insoluble portion of PE-*v* samples also display a low-*q* upturn that follows the same power law as the neat and peroxide crosslinked PE samples. Because this liquid-like scattering does not seem to provide relevant information regarding the PE-*g* and PE-*v* nanostructure, its contribution was removed from the PE-*g*, PE-*v*, and insoluble portion of PE-*v* patterns using the technique developed by Roe and Gieniewski.[S5] After the initial subtraction of the empty sample cell scattering, a linear regression was performed on the low- and high-*q* regions of the SAXS patterns (*i.e.*, $q < 0.006$ nm$^{-1}$ and $q > 3.7$ nm$^{-1}$) using the equation

$$Iq^{3.15} = mq^{3.15} + b \qquad (S3)$$

where $I$ is the SAXS pattern intensity, $m$ is the regression slope, and $b$ is the regression intercept. Figures S11B and S11C feature the background regression for the PE-*g*-4 SAXS pattern. The final SAXS pattern was produced by subtracting the calculated background contribution (Figure S11D). To avoid the influence of background subtraction artifacts, only the region between $0.01$ nm$^{-1} \leq q \leq 1$ nm$^{-1}$ was used in subsequent SAXS pattern analyses.

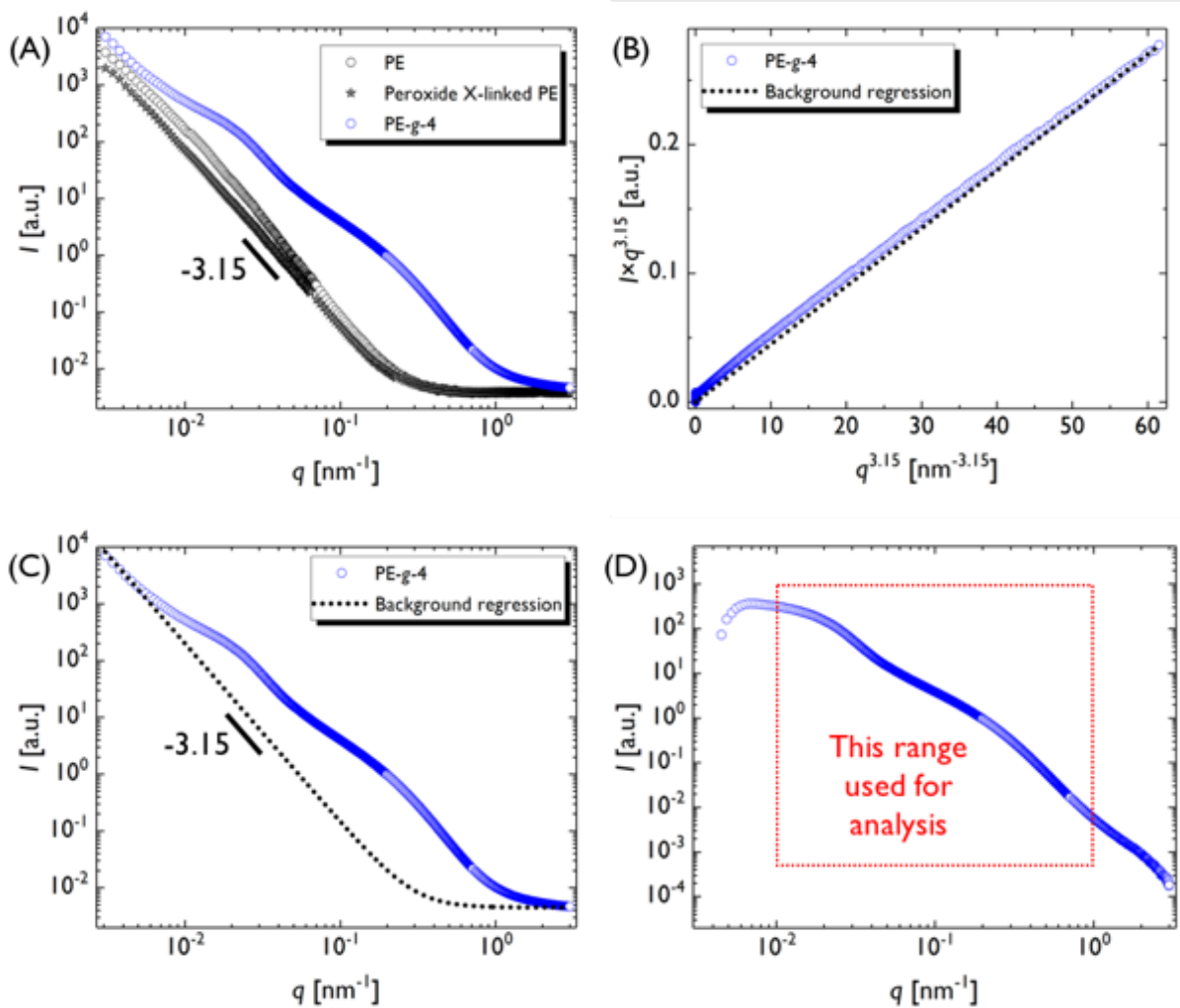

**Figure S11.** (A) Non-background-subtracted SAXS patterns of PE, peroxide crosslinked PE, and PE-*g*-4 at 160 °C. (B & C) Comparison between PE-*g*-4 SAXS pattern and background regression. (D) Background-subtracted SAXS pattern of PE-*g*-4.

*Comparison of initial PE-v and insoluble portion of PE-v SAXS patterns*

Figure S12 contains the SAXS patterns of PE-*v* with varying crosslink density and insoluble portion of PE-*v* samples. Figure S13 compares the SAXS patterns of the initial PE-*v* samples against their corresponding insoluble portion of PE-*v* samples. Figure S14 shows that the vertically-shifted SAXS pattern of each initial PE-*v* sample overlaps the SAXS pattern of its analogous insoluble portion.

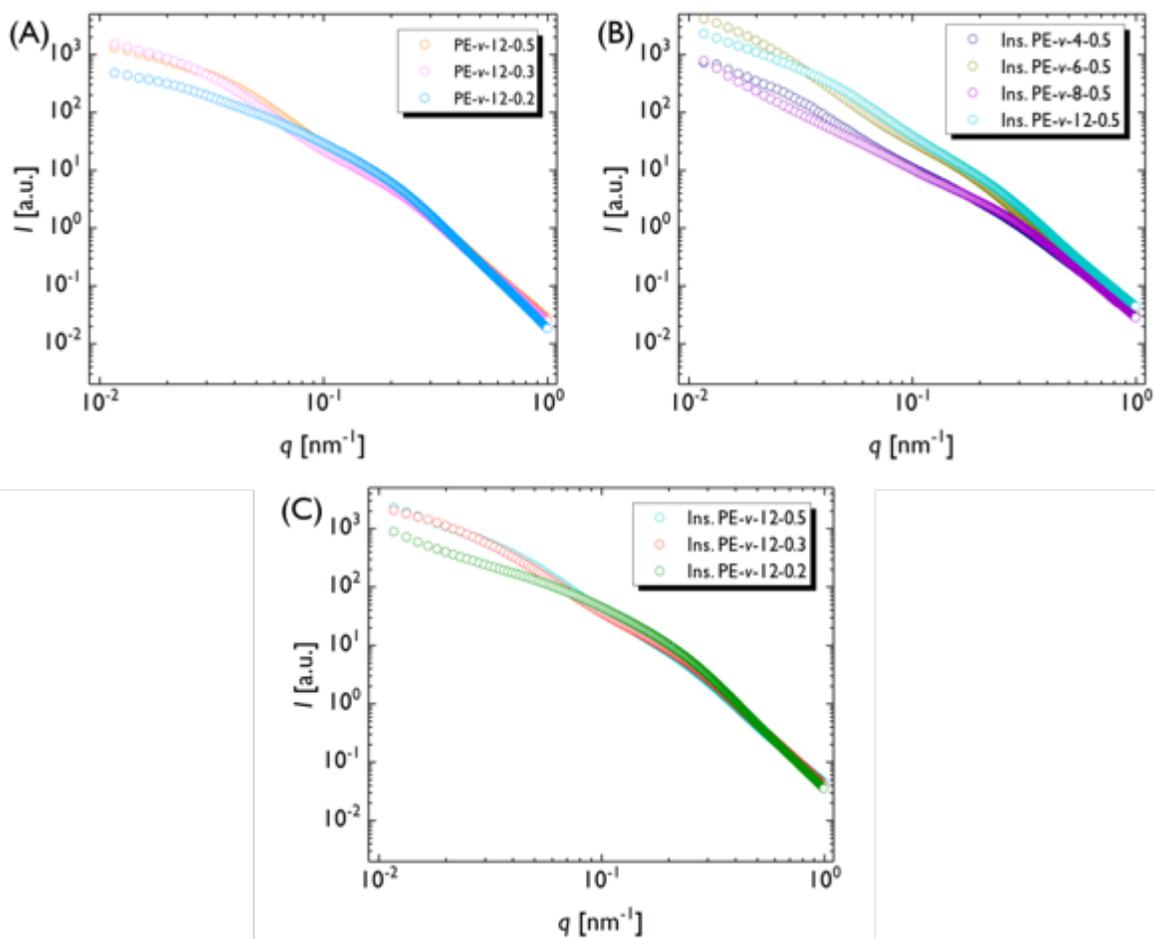

**Figure S12.** SAXS patterns at 160 °C of (A) PE-*v* with varying crosslink density, (B) insoluble portion of PE-*v* samples with varying graft density, and (C) insoluble portion of PE-*v* samples with varying crosslink density. The insoluble portion of PE-*v*-12-0.5 patterns in (B) and (C) are the same.

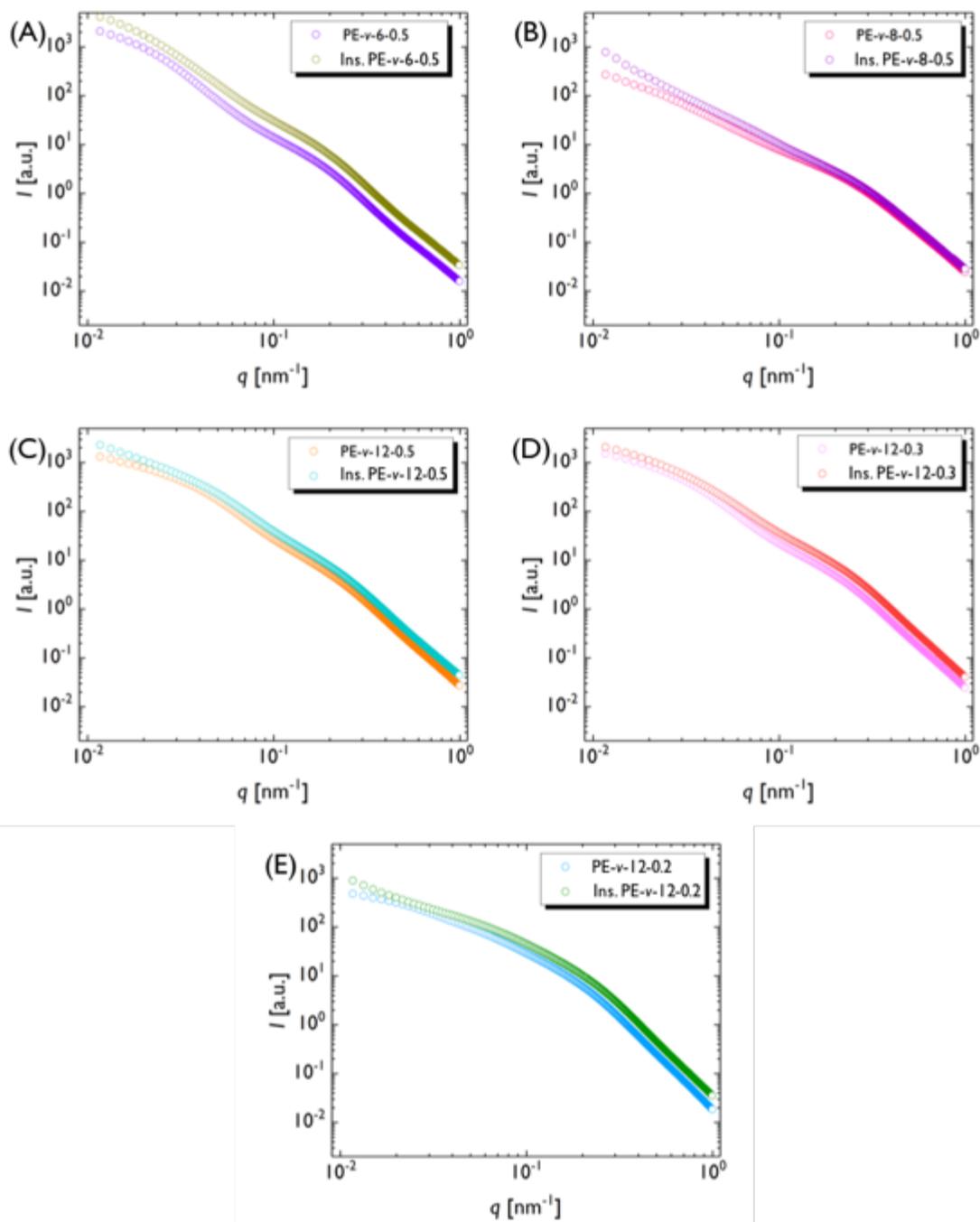

**Figure S13.** Comparison of initial and insoluble portion of vitrimer SAXS patterns at 160 °C for (A) PE-*v*-6-0.5, (B) PE-*v*-8-0.5, (C) PE-*v*-12-0.5, (D) PE-*v*-12-0.3, and (E) PE-*v*-12-0.2.

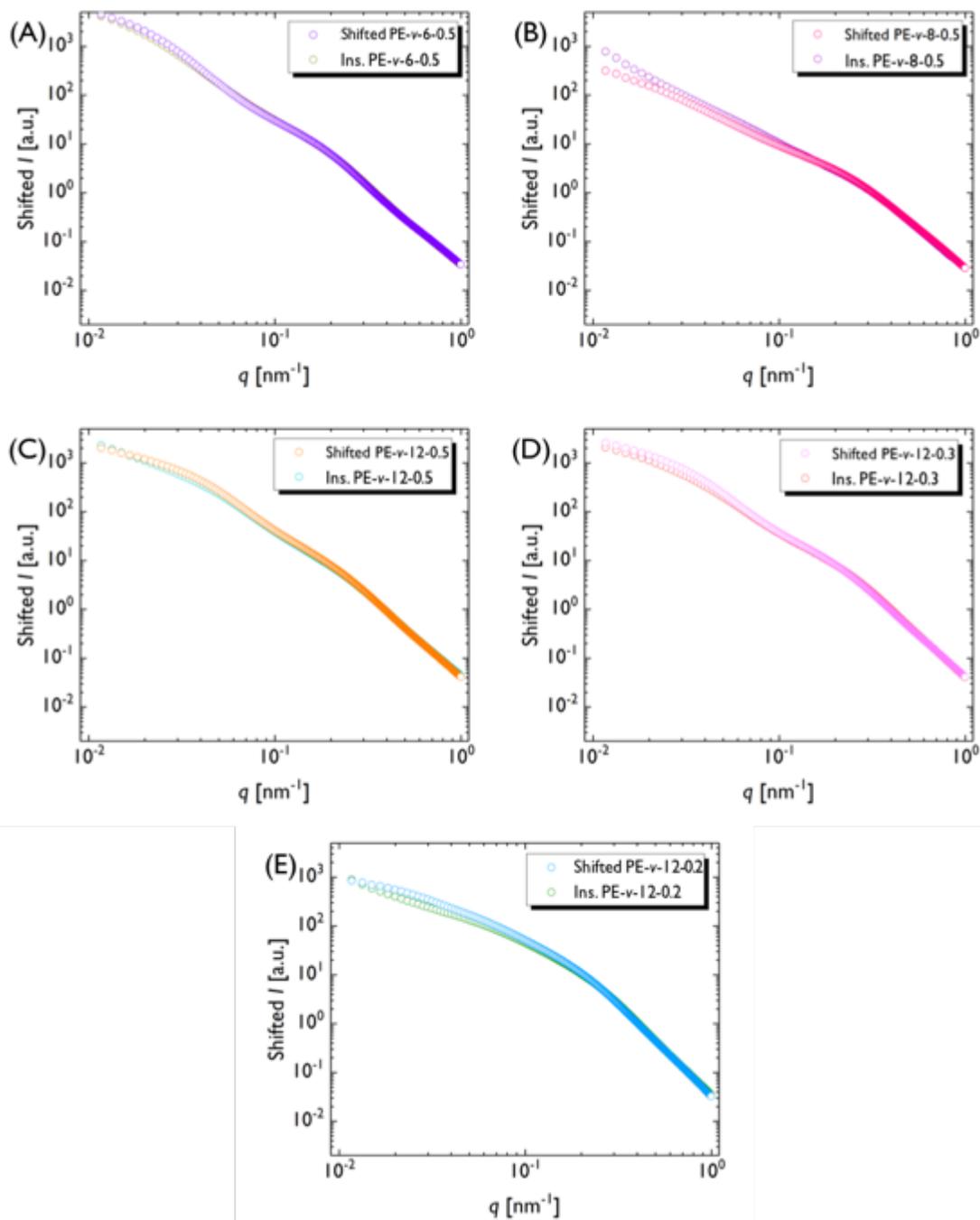

**Figure S14.** Comparison of shifted initial vitrimer and non-shifted insoluble portion of vitrimer SAXS patterns at 160 °C for (A) PE-*v*-6-0.5, (B) PE-*v*-8-0.5, (C) PE-*v*-12-0.5, (D) PE-*v*-12-0.3, and (E) PE-*v*-12-0.2.

*Influence of annealing time and temperature on SAXS patterns*

The nanostructure of molten PE-*v*-12-0.5 under varying conditions was examined using synchrotron-sourced SAXS at both beamline SWING at the SOLEIL Synchrotron Source (Saint-Aubin, FR) and the DND-CAT beamline 5-ID-D at the Advanced Photon Source in Argonne National Laboratory (Argonne, IL). For measurements performed at SWING, the sample-to-detector distance was 6.1 m, the X-ray wavelength was 1.03 Å, and the 2D scattering patterns were collected by a Eiger 4M CCD detector. For experiments performed at DND-CAT, the sample-to-detector distance was 8.5 m, the X-ray wavelength was 1.03 Å, and the 2D scattering patterns were collected by a Rayonix CCD area detector. To probe the influence of annealing time, PE-*v*-12-0.5 was initially heated at 160 °C for 5 min inside a Linkam hot stage placed in the X-ray beam path. After collecting a SAXS pattern, the sample was removed from the hot stage and immediately placed onto a heating plate pre-heated to 160 °C. The sample was annealed on the heating plate for 48 hrs, and then immediately transferred into the hot stage pre-heated to 160 °C. The transfer time was less than 5 seconds. Figure S15A shows that despite the large difference between annealing times, the SAXS patterns collected at each time point were identical. Figures 15B also shows that temperature had no influence on the SAXS pattern of molten PE-*v*-12-0.5. These results suggest that the observed nanostructures of molten PE-*g* and PE-*v* are stable for relatively long experimental time scales. Further experiments are needed to determine if these nanostructures are kinetically trapped or at thermodynamic equilibrium.

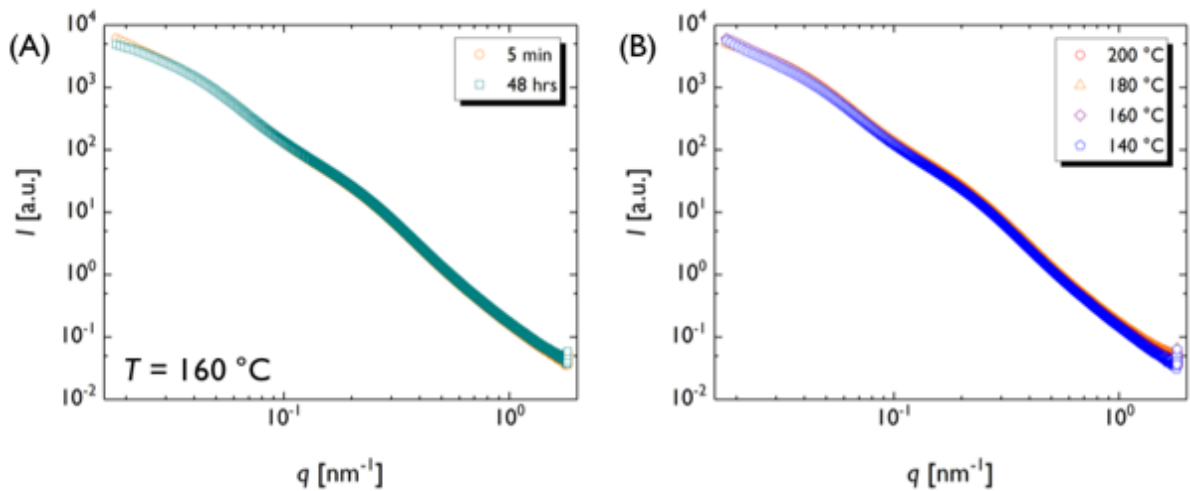

**Figure S15.** Comparison of molten PE-*v*-12-0.5 SAXS patterns at varying (A) annealing time and (B) temperature. Note: the background contribution was not removed from these patterns.

*Comparison of SAXS patterns at 40 and 160 °C*

Figures S16–S18 are the 40 and 160 °C SAXS patterns of the PE-*g*, PE-*v*, and insoluble portion of PE-*v* samples. The 160 °C patterns were vertically shifted to show that they overlap with the low-*q* region of the 40 °C patterns.

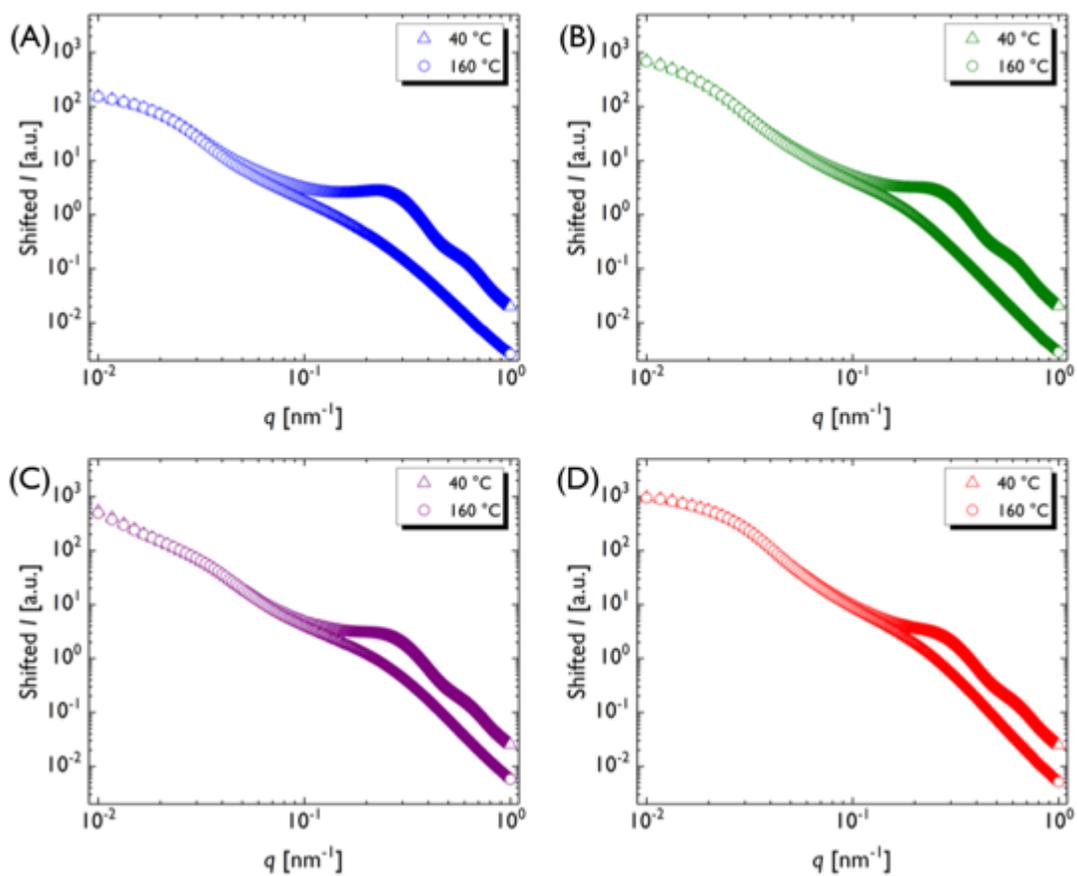

**Figure S16.** Comparison of shifted PE-*g* SAXS patterns at 40 and 160 °C. (A) PE-*g*-4, (B) PE-*g*-6, (C) PE-*g*-8, and (D) PE-*g*-12.

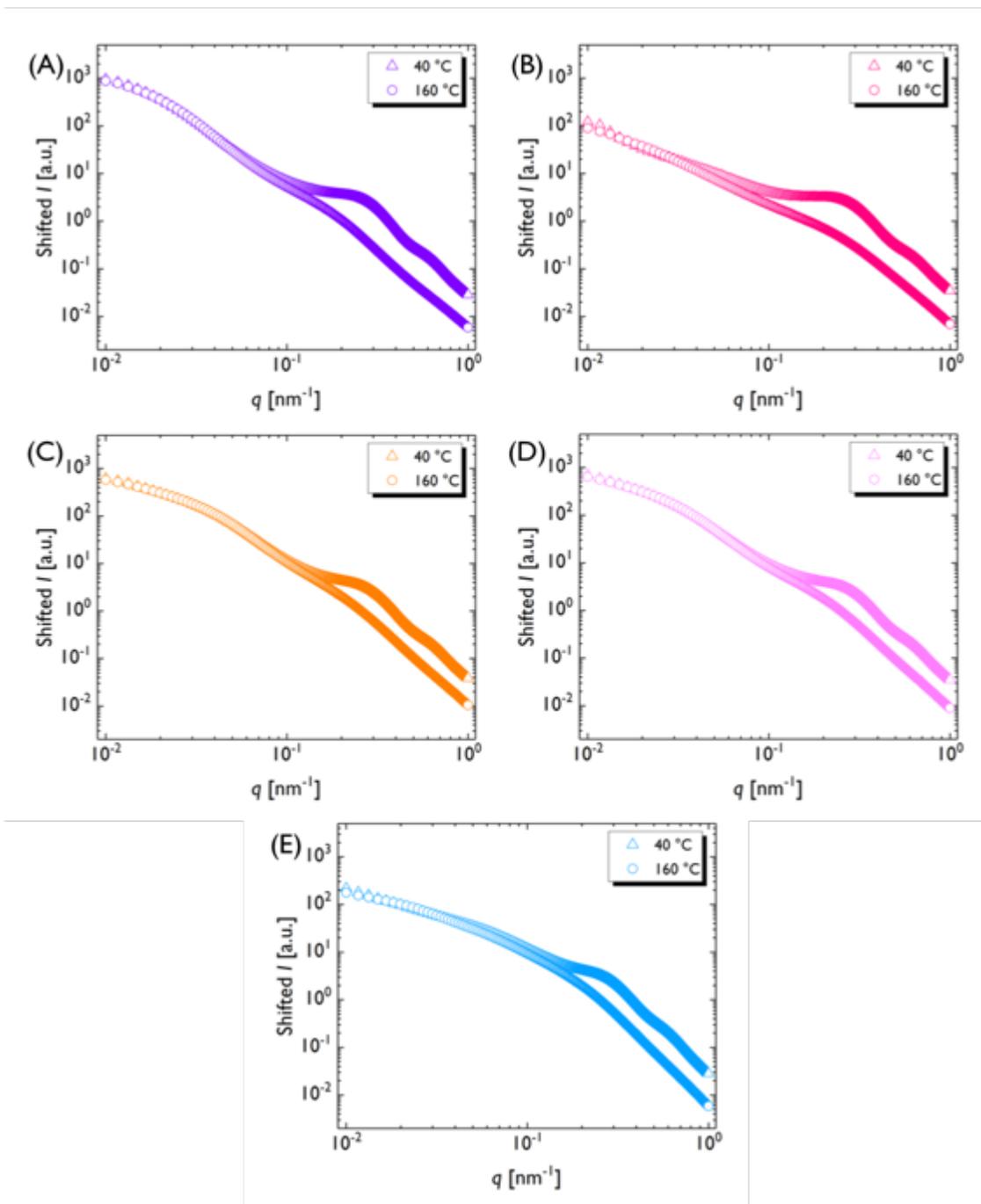

**Figure S17.** Comparison of shifted PE-*v* SAXS patterns at 40 and 160 °C. (A) PE-*v*-6-0.5, (B) PE-*v*-8-0.5, (C) PE-*v*-12-0.5, (D) PE-*v*-12-0.3, and (E) PE-*v*-12-0.2.

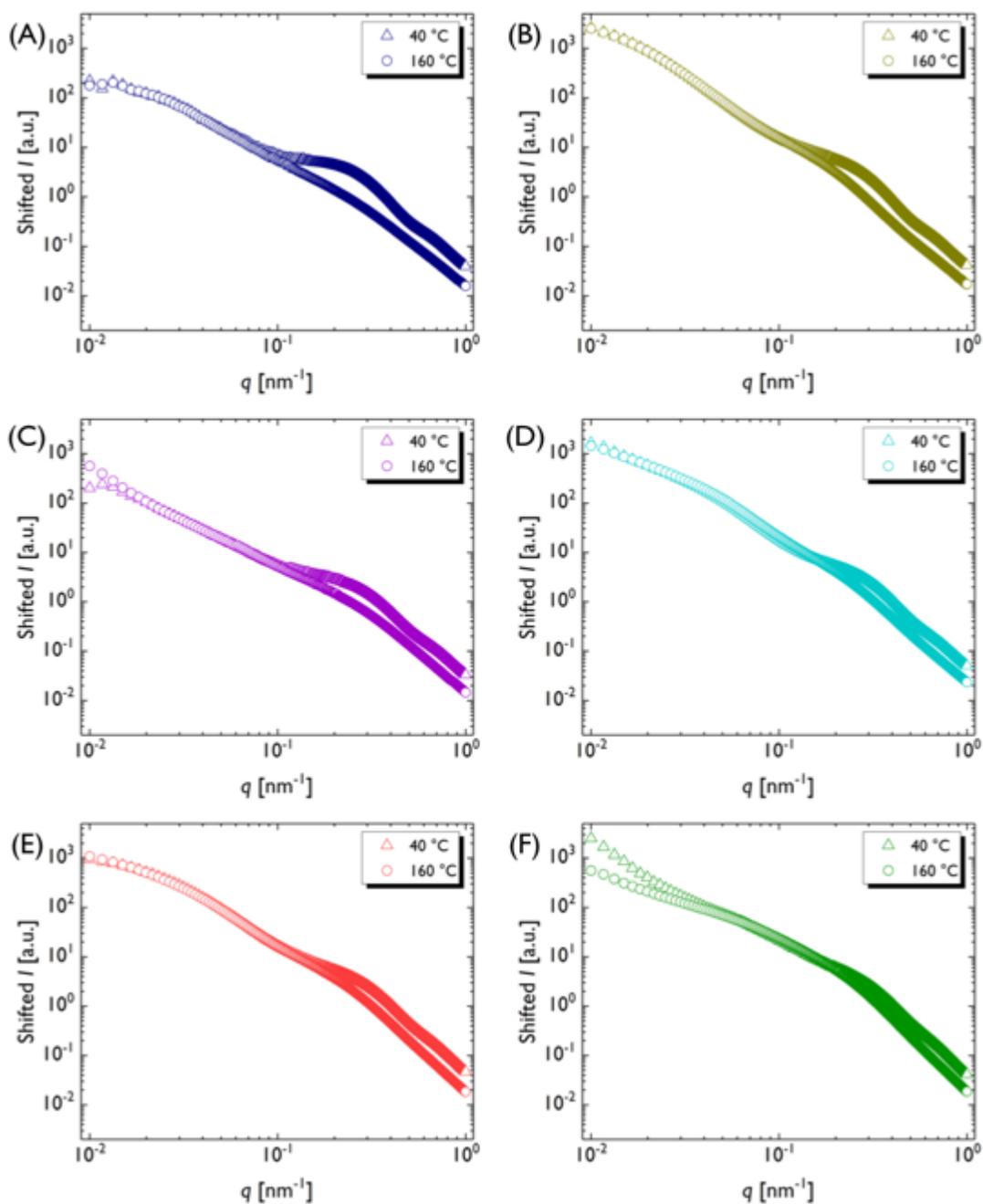

**Figure S18.** Comparison of shifted insoluble portion of PE-*v* SAXS patterns at 40 and 160 °C. (A) Insoluble portion of PE-*v*-4-0.5, (B) insoluble portion of PE-*v*-6-0.5, (C) insoluble portion of PE-*v*-8-0.5, (D) insoluble portion of PE-*v*-12-0.5, (E) insoluble portion of PE-*v*-12-0.3, and (F) insoluble portion of PE-*v*-12-0.2.

*Wide-angle X-ray scattering patterns*

Synchrotron-sourced wide-angle X-ray scattering (WAXS) patterns for PE, peroxide crosslinked PE, PE-*g*, PE-*v*, and insoluble portion of PE-*v* samples are featured in Figure S19. All WAXS patterns display the characteristic Bragg reflections of the PE orthorhombic unit cell with space group *Pnam* (see Table S2 for peak indexing). From the WAXS patterns, the unit cell parameters $a$, $b$, and $c$ were measured using the GSAS-II Crystallography software package.[S6] The measured unit cell parameters are approximately the same for all samples (Figure S20), suggesting that the presence of the grafts and crosslinks – at the densities examined in this work – have minimal influence on the PE unit cell dimensions.

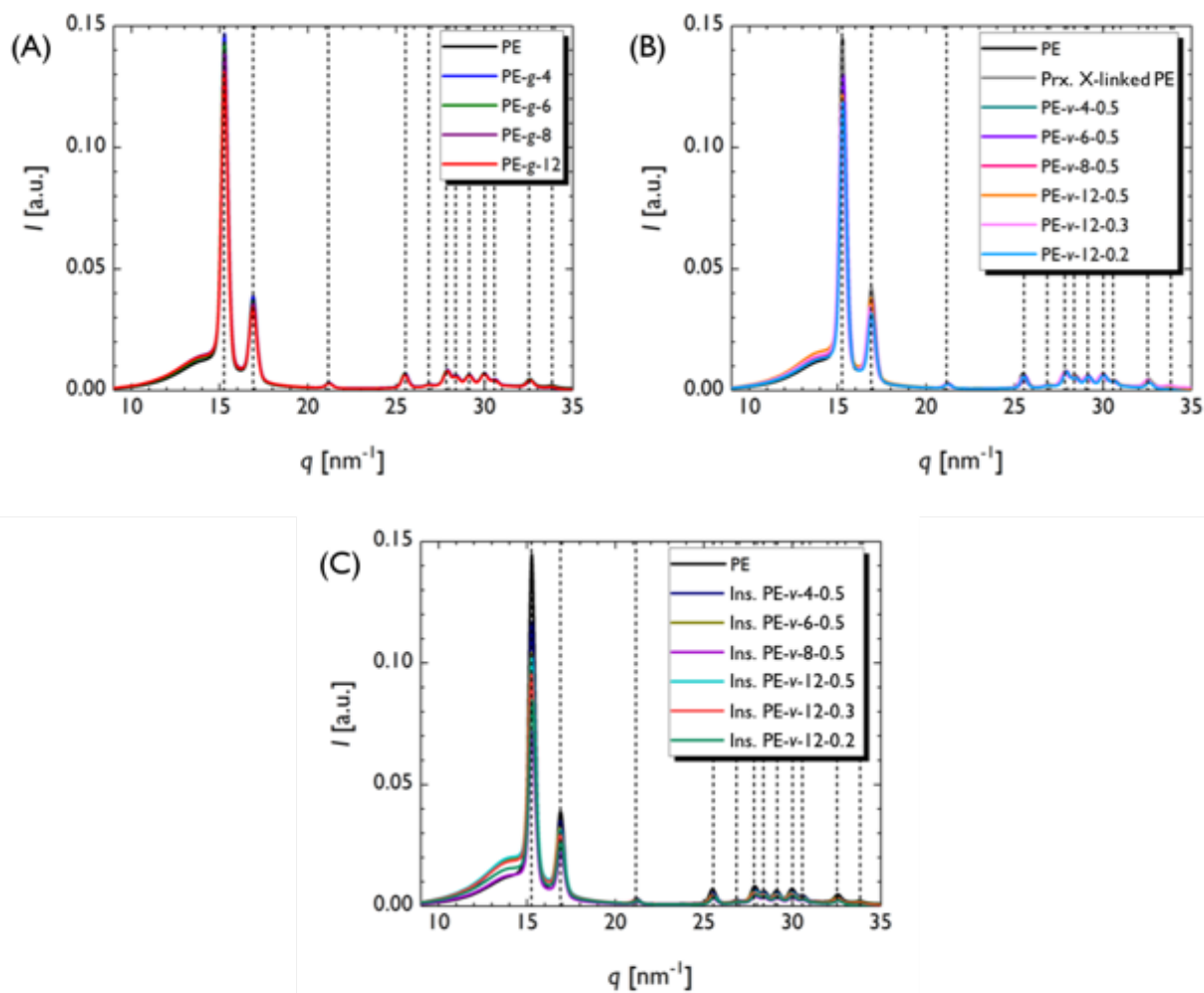

**Figure S19.** WAXS patterns of (A) PE-*g*, (B) PE-*v* and peroxide crosslinked (Prx. X-linked) PE, and (C) the insoluble portion of PE-*v*. The PE WAXS pattern is displayed in all patterns for comparison, while the vertical dashed lines signify characteristic reflections of the orthorhombic PE crystal unit cell with space group *Pnam*.

**Table S2.** PE orthorhombic unit cell (space group *Pnam*) Miller indices corresponding to observed WAXS Bragg peaks.

| $q$ [nm$^{-1}$] | [h k l] |
|---|---|
| 15.3 | 1 1 0 |
| 16.9 | 2 0 0 |
| 21.2 | 2 1 0 |
| 25.5 | 0 2 0 |
| 26.8 | 1 2 0 |
| 27.8 | 0 1 1 |
| 28.3 | 3 1 0 |
| 29.1 | 1 1 1 |
| 30.0 | 2 0 1 |
| 30.6 | 2 2 0 |
| 32.5 | 2 1 1 |
| 33.8 | 4 0 0 |

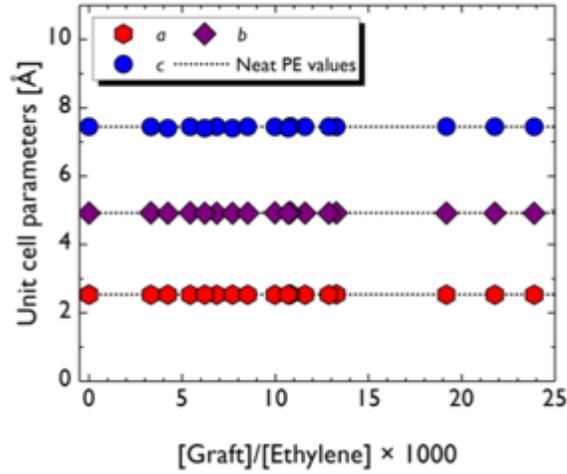

**Figure S20.** Measured orthorhombic unit cell dimensions (space group *Pnam*) for PE, peroxide crosslinked PE, PE-*g*, PE-*v*, and insoluble portion of PE-*v*.

*PE crystallite dimensions quantification via electron density correlation function analysis*

The dimensions of the PE crystallite lamellae may be quantified by converting the SAXS patterns of semi-crystalline PE, PE-*g*, and PE-*v* into electron density correlation functions.[S7,S8,S9] For a two-phase system composed of isotropically distributed stacks of parallel lamellae, the intensity distribution, $\Sigma(q)$, is described by

$$\Sigma(q) = \frac{2\pi}{q^2} r_e^2 \int_{-\infty}^{\infty} \exp(-iqr) P(r) dr \tag{S4}$$

where $q$ is the scattering vector, $r_e$ is the electron radius, $r$ is the distance along the trajectory that is normal to the surface of a lamellae stack, and $P(r)$ is the one-dimensional electron density correlation function. The inverse Fourier transform of Equation S4 gives

$$P(r) = \frac{1}{2\pi^2 r_e^2} \int_{-\infty}^{\infty} \exp(iqr) q^2 \Sigma(q) \, dq = \frac{1}{2\pi^2 r_e^2} \int_{0}^{\infty} \cos(qr) \, q^2 \Sigma(q) \, dq \tag{S5}$$

If the two-phase layer system is not perfectly ordered (*i.e.*, fluctuations of intercrystalline spacings, varying lamellae thicknesses, and diffuse interfaces), $P(r)$ exhibits a characteristic shape that is similar to a damped sine wave.[7] Various features of the characteristic shape correspond to the average amorphous layer thickness ($L_a$) and the long period spacing ($L_p$). The average lamella thickness ($L_c$) is the difference between $L_p$ and $L_a$.

The arbitrary intensity units of the SAXS patterns reported in this manuscript prevent direct calculation of $P(r)$. Instead, each semi-crystalline PE-g, PE-v, and insoluble portion of PE-v SAXS pattern was converted into a non-absolute unit electron density correlation function, $K(r)$, using the equation

$$K(r) = \int_0^\infty \cos(qr)\, q^2 I(q)\, dq \tag{S6}$$

where $I(q)$ is the SAXS scattering pattern intensity in arbitrary units. Despite the use of arbitrary units, $K(r)$ has the same characteristic shape and horizontal positioning as $P(r)$, so determination $L_a$, $L_p$, and $L_c$ is the same.

Figure S21A displays the estimated $K(r)$ curves for neat PE. The curve peak corresponded to an $L_p$ value of $21 \pm 1$ nm. $L_a$ was found by finding the intersection between the $K(r)$ minimum and $dK/dr$, which is an extrapolation of the linear portion of the curve near $r = 0$. $L_a$ was estimated to be $4 \pm 1$ nm. Accordingly, $L_c$ was $16 \pm 2$ nm.

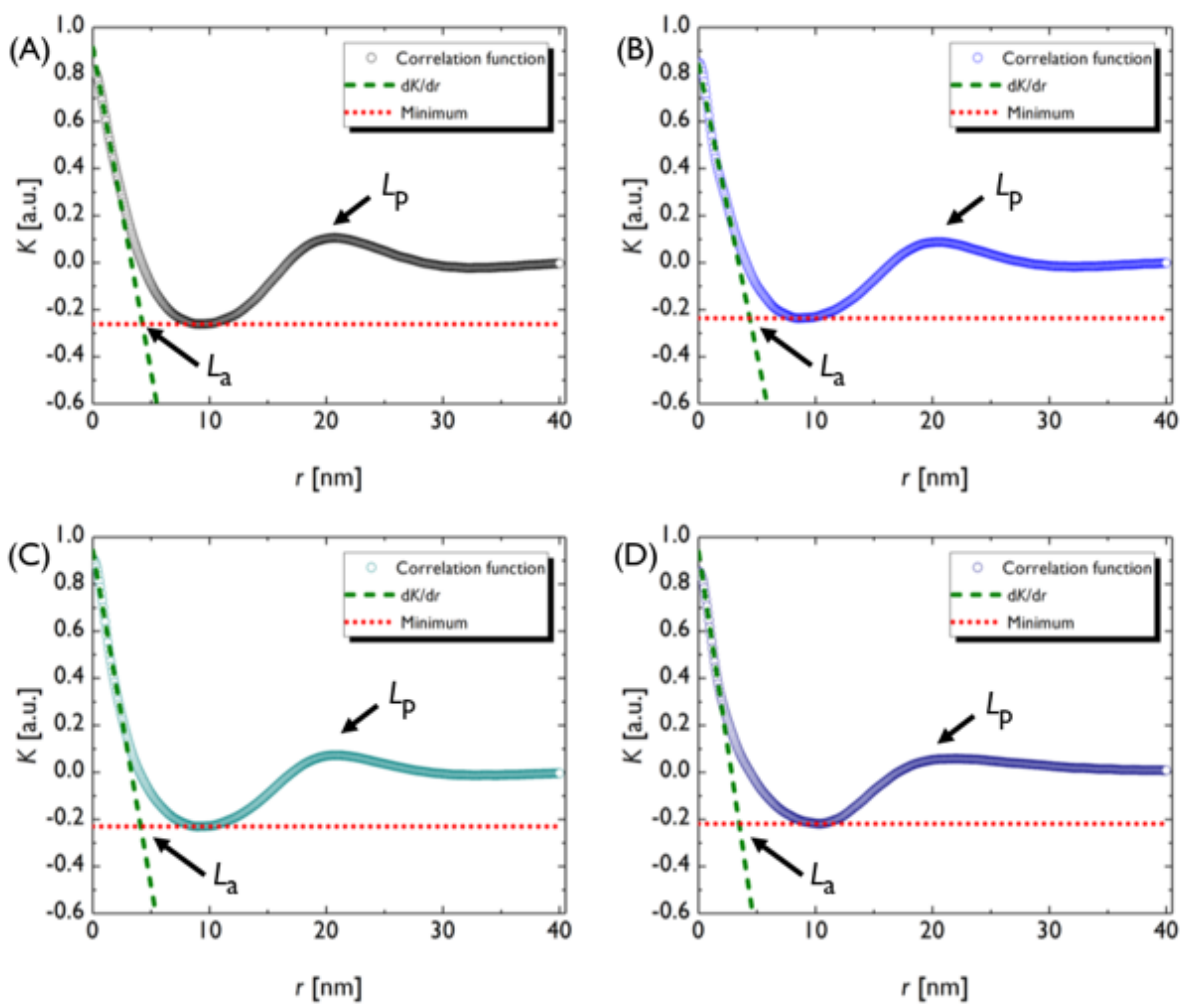

**Figure S21.** Non-absolute unit electron density correlation functions for (A) PE, (B) PE-*g*-4, (C) PE-*v*-4-0.5, and (D) insoluble portion of PE-*v*-4-0.5.

To satisfy the two-state model assumption for the PE-*g*, PE-*v*, and insoluble portion of PE-*v* SAXS patterns, the scattering contribution of the amorphous nanostructure was first removed. This was achieved by subtracting the shifted melt pattern from the semi-crystalline SAXS pattern. The resulting $K(r)$ curves and estimated crystallite dimensions are displayed in Figures S21B–21D and S22A, respectively. The local degree of crystallinity, $X_K$, is given by the quotient of $L_c/L_p$. For all samples, $X_K$ remains around a value of 0.80 (Figure S22B), which contrasts the values and trend for the total crystallinity measured by DSC. This discrepancy – also seen in other systems – is due to the fact that $X_K$ only describes the crystallinity within the lamellar stacks, while the crystallinity measured by DSC characterizes the entire sample.[8] To convert $X_K$ to a mass fraction basis, crystalline and amorphous PE densities of 1.000 and 0.853 g/mL, respectively, were used.[S10] For the graft and crosslinker, a density of 1.4 g/mL – a typical value for small molecules – was assumed.[S11]

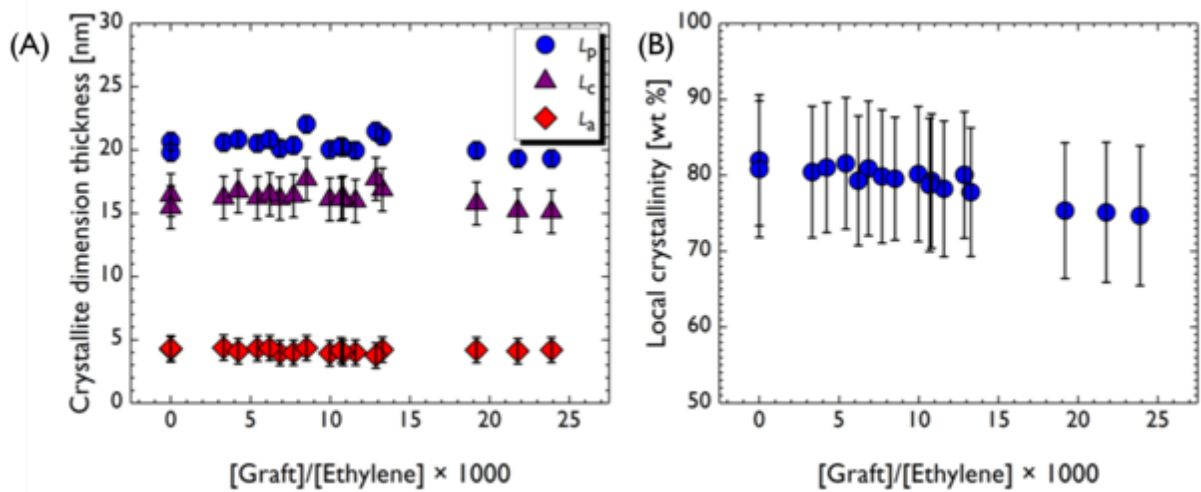

**Figure S22.** (A) Crystallite dimension thicknesses for PE, peroxide crosslinked PE, PE-*g*, PE-*v*, and insoluble portion of PE-*v*. Error bars for $L_p$, $L_a$, and $L_c$ represent the standard deviation of all samples, the range of estimated values, and the propagated error, respectively. (B) Estimated local crystallinity values. Error bars represent the propagated error.

*Polarized optical microscopy*

Polarized optical microscopy was conducted with a LEICA DMR DAS microscope equipped with a 10×/0.30 HC PL Fluotar objective lens. A 150 μm thick film of the sample was sandwiched between a glass microscope slide and coverslip, and then put into a Linkam Scientific LTS350 heating stage. Prior to each image collection, the sample was annealed at 160 °C for 5 minutes, then cooled to 40 °C at a rate of -10 °C/min to induce crystallization. Figure S23 features polarized optical microscopy images of neat PE, peroxide crosslinked PE, PE-*g*-4, PE-*v*-4-0.5, and the insoluble portion of PE-*v*-4-0.5. While neat and peroxide crosslinked PE show the characteristic signs of spherulitic growth (*i.e.*, Maltese cross and optical banding patterns),[S12] PE-*g*-4, PE-*v*-4-0.5, and the insoluble portion of PE-*v*-4-0.5 show no evidence of spherulite structures.

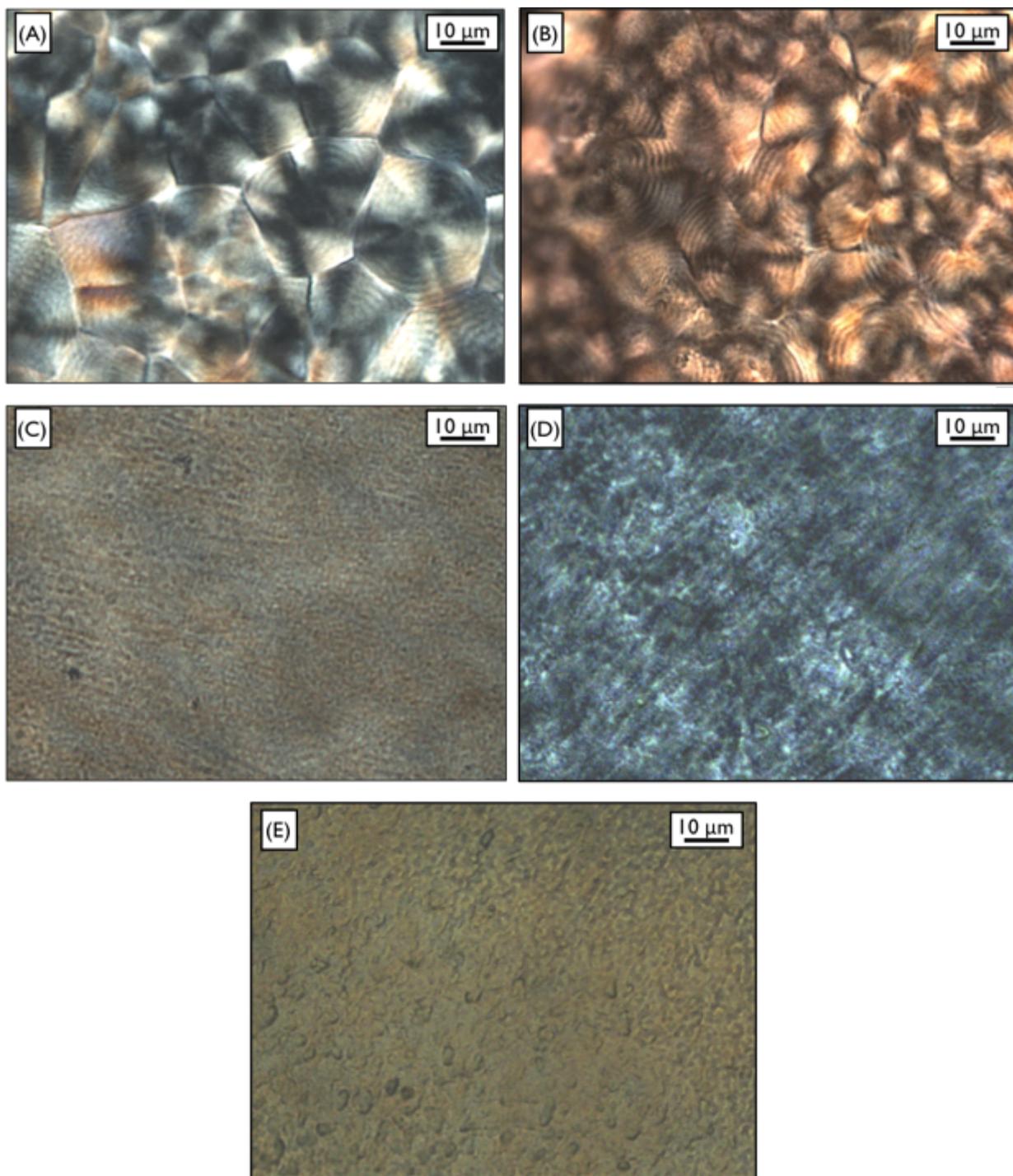

**Figure S23.** Polarized optical microscopy images of (A) PE, (B) peroxide crosslinked PE, (C) PE-*g*-4, (D) PE-*v*-4-0.5, and (E) insoluble portion of PE-*v*-4-0.5.

*Hypothesized semi-crystalline nanostructure for PE-g and PE-v*

Figure S24 illustrates the hypothesized semi-crystalline nanostructure for PE-*g* and PE-*v* materials. It is posited that nucleation and growth of PE crystal lamellae confine the graft-rich aggregates into the amorphous layers.

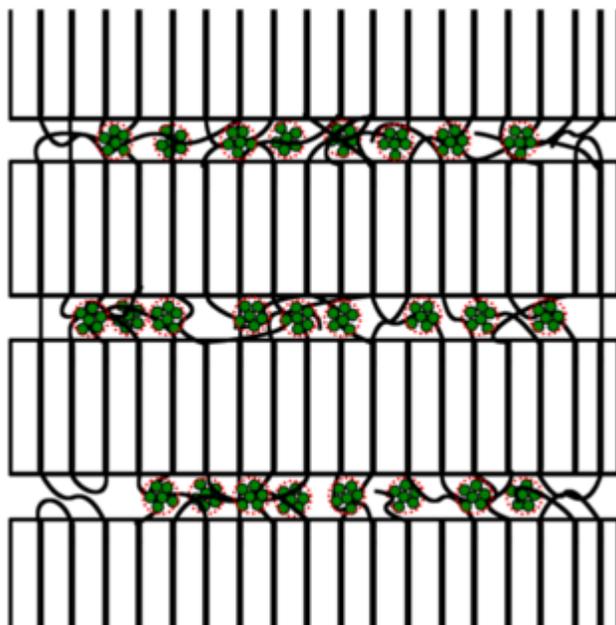

**Figure S24.** Hypothesized nanostructure of semi-crystalline PE-*g* and PE-*v* materials.

*SAXS pattern fits to the 2 term Debye-Bueche scattering model*

In addition to the aggregate-fractal scattering model, SAXS patterns were also fit to the following empirical 2 term Debye-Bueche scattering model:

$$I(q) = \frac{\alpha_{low}\xi_{low}^3}{(1+(q\xi_{low})^2)^2} + \frac{\alpha_{high}\xi_{high}^3}{(1+(q\xi_{high})^2)^2} \tag{S7}$$

where $I$ is the SAXS pattern intensity, $q$ is the scattering vector, $\alpha_{low}$ and $\alpha_{high}$ are the amplitude factors of the low- and high-$q$ shoulders, respectively, and $\xi_{low}$ and $\xi_{high}$ are the corresponding length scales associated with those features. Model fitting was performed using custom-made

MATLAB (version R2017b) scripts. Although both terms in this model are generic scattering equations,[S13] the model describes the measured SAXS patterns very well (Figure S25A). Table S3 lists the estimated values for $\alpha_{low}$, $\xi_{low}$, $\alpha_{high}$, and $\xi_{high}$. The linear relationship between $\alpha_{high}$ and graft density suggests that the high-$q$ shoulder scattering intensity is directly proportional to the volume fraction of some minority component in the sample (Figure S25B).

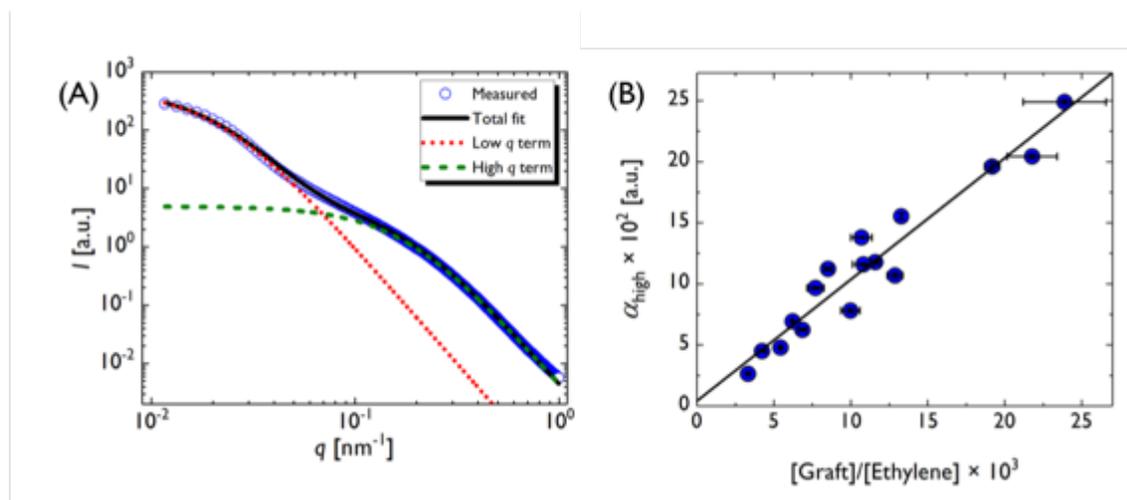

**Figure S25.** (A) SAXS pattern of PE-$g$-4 fit to the 2 term Debye-Bueche scattering model. (B) Estimated $\alpha_{high}$ for PE-$g$, PE-$v$, and insoluble portion of PE-$v$ samples. Vertical error bars are the standard error of the nonlinear regression.

**Table S3.** Estimated parameters from 2 term Debye-Bueche scattering model fittings. Error bars represent the standard error of the nonlinear regression.

| Sample | $\alpha_{low} \times 10^2$ [a.u.] | $\xi_{low}$ [nm] | $\alpha_{high} \times 10^2$ [a.u.] | $\xi_{high}$ [nm] |
|---|---|---|---|---|
| PE-$g$-4 | 0.5 ± 0.1 | 47 ± 1 | 2.6 ± 0.1 | 6 ± 1 |
| PE-$g$-6 | 1.4 ± 0.1 | 61 ± 1 | 4.8 ± 0.1 | 8 ± 1 |
| PE-$g$-8 | 1.2 ± 0.1 | 31 ± 1 | 6.2 ± 0.1 | 5 ± 1 |
| PE-$g$-12 | 5.2 ± 0.1 | 39 ± 1 | 7.8 ± 0.1 | 7 ± 1 |
| PE-$v$-4-0.5 | 0.7 ± 0.1 | 39 ± 1 | 4.5 ± 0.1 | 4 ± 1 |
| PE-$v$-6-0.5 | 3.0 ± 0.1 | 50 ± 1 | 6.9 ± 0.1 | 6 ± 1 |
| PE-$v$-8-0.5 | 0.7 ± 0.1 | 36 ± 1 | 9.7 ± 0.1 | 4 ± 1 |
| PE-$v$-12-0.5 | 5.6 ± 0.1 | 30 ± 1 | 11.8 ± 0.1 | 6 ± 1 |
| PE-$v$-12-0.3 | 4.5 ± 0.1 | 35 ± 1 | 11.6 ± 0.1 | 5 ± 1 |
| PE-$v$-12-0.2 | 1.7 ± 0.1 | 31 ± 1 | 13.8 ± 0.1 | 8 ± 1 |
| Insoluble portion of PE-$v$-4-0.5 | 1.5 ± 0.1 | 41 ± 1 | 11.2 ± 0.1 | 5 ± 1 |
| Insoluble portion of PE-$v$-6-0.5 | 5.4 ± 0.1 | 52 ± 1 | 15.5 ± 0.2 | 6 ± 1 |
| Insoluble portion of PE-$v$-8-0.5 | 0.7 ± 0.1 | 60 ± 1 | 10.7 ± 0.2 | 5 ± 1 |
| Insoluble portion of PE-$v$-12-0.5 | 6.4 ± 0.1 | 34 ± 1 | 19.6 ± 0.4 | 6 ± 1 |
| Insoluble portion of PE-$v$-12-0.3 | 5.8 ± 0.1 | 35 ± 1 | 20.4 ± 0.2 | 6 ± 1 |
| Insoluble portion of PE-$v$-12-0.2 | 1.4 ± 0.1 | 41 ± 1 | 24 ± 0.2 | 8 ± 1 |

*SAXS pattern fits to the aggregate-fractal scattering model*

SAXS patterns of PE-*g*, PE-*v*, and insoluble portion of PE-*v* at 160 °C were fit to the aggregate-fractal scattering model detailed in the main text. In this model, scattering from the individual aggregates is described using a polydisperse spheres form factor ($P_{sph}$) with the following functional form:

$$P_{sph}(q) = \int_0^\infty f(r)\Phi^2(qr)dr \tag{S8}$$

$$\Phi(x) = \frac{3[\sin(x) - x\cos(x)]}{x^3} \tag{S9}$$

$$f(r) = \frac{z^{z+1}}{\Gamma(z+1)} \frac{r^{z-1}}{R_{sph}^z} \exp\left(\frac{-zr}{R_{sph}}\right) \tag{S10}$$

$$z = \frac{1}{\sigma^2} - 1 \tag{S11}$$

where $f$ is the normalized Schulz-Zimm distribution,[S14] $r$ is the aggregate radius, $q$ is the scattering vector, $\Phi$ represents the square root of the hard sphere form factor, $R_{sph}$ is the average aggregate radius, $z$ is the Schulz-Zimm distribution width parameter, and $\sigma$ is the standard deviation of the aggregate radius. To fit the data to the model, $\alpha_{sph}$, the amplitude factor for scattering from an individual aggregate, was assumed to be linearly proportional to the graft density. Based on this assumption, a calibration curve was developed to enable calculation of $\alpha_{sph}$ *a priori*. Specifically, preliminary model fits – in which $R_{sph}$ was set to 2 nm and the parameters $\alpha_{sph}$, $\sigma$, $\alpha_{frac}$, $D$, and $\xi$ freely floated – were performed on the PE-*g* SAXS patterns. A linear regression was performed between the estimated $\alpha_{sph}$ and grafting density. The resulting linear equation was then used to calculate the $\alpha_{sph}$ values used in the final SAXS pattern fittings (Figure S26). Figure S27 displays the model fit for the PE-g, PE-v, and insoluble portion of PE-v samples. Figures S28 and S29 detail the fitted values for $\sigma$, $\alpha_{frac}$, and $D$.

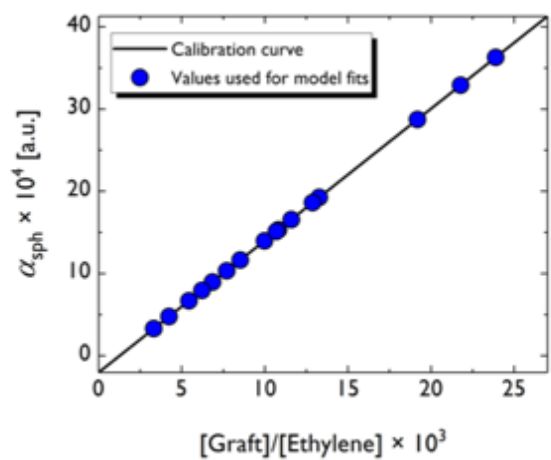

**Figure S26.** Calibration curve for the $\alpha_{sph}$ parameter in aggregate-fractal scattering model.

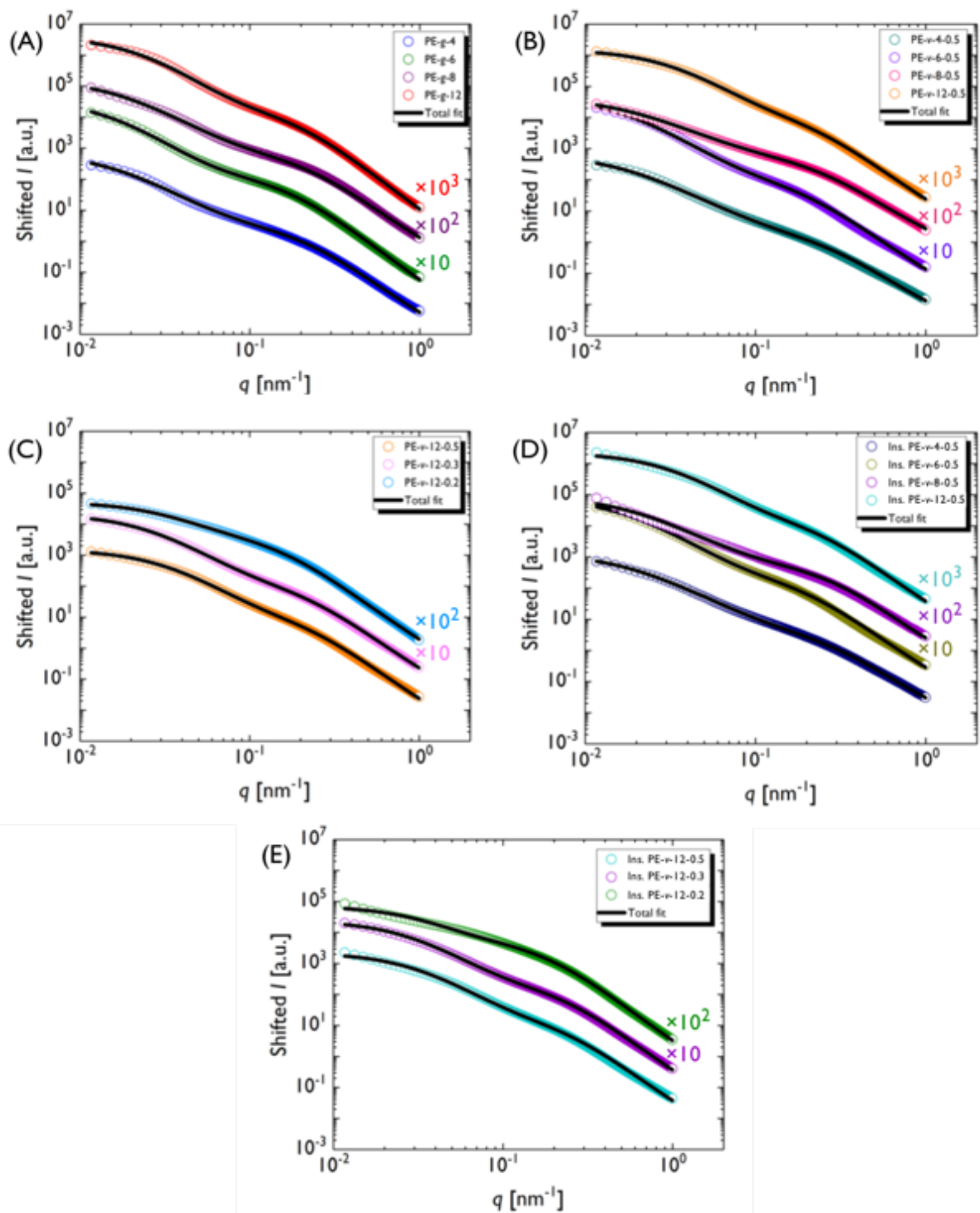

**Figure S27.** SAXS patterns at 160 °C fit to the aggregate-fractal scattering model. (A) PE-*g* with varying graft density, (B) PE-*v* with varying graft density, (C) PE-*v* with varying crosslink density, (D) insoluble portion of PE-*v* samples with varying graft density, and (E) insoluble portion of PE-*v* samples with varying crosslink density. The PE-*v*-12-0.5 patterns in (B) and (C) are the same, while the insoluble portion of PE-*v*-12-0.5 patterns in (D) and (E) are the same. SAXS patterns were shifted for clarity.

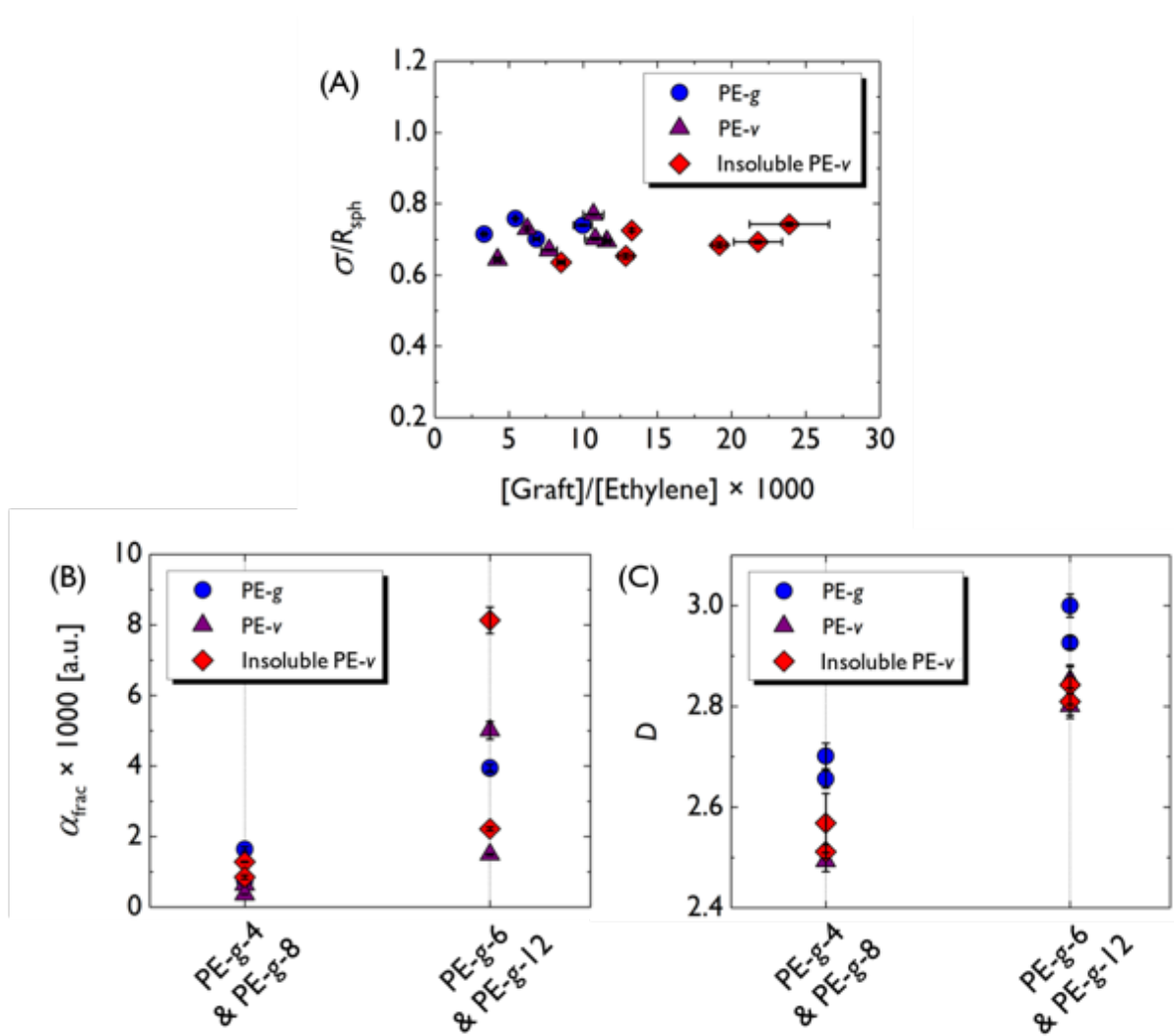

**Figure S28.** Estimated aggregate-fractal scattering model fit parameters. (A) Aggregate radius dispersity, (B) amplitude factor for scattering from a mass fractal for samples with varying graft density, and (C) fractal dimension for samples with varying graft density. Error bars are the standard error of the nonlinear regression. In (B) and (C), each PE-*v* and insoluble portion of PE-*v* sample is grouped with its corresponding PE-*g* sample (*e.g.*, PE-*v*-4-0.5 and PE-*v*-8-0.5 are listed in the PE-*g*-4 & PE-*g*-8 group.)

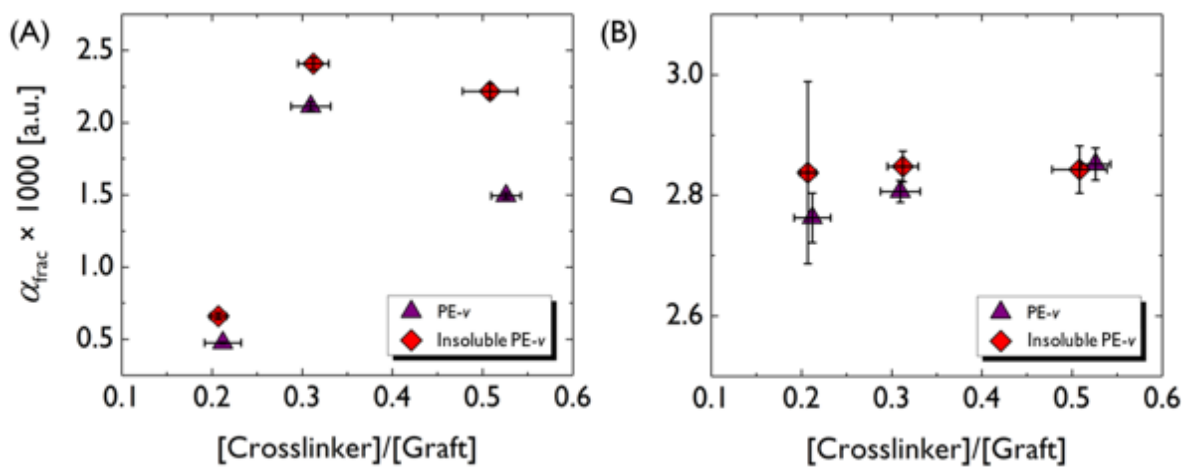

**Figure S29.** Estimated aggregate-fractal scattering model fit parameters for initial and insoluble portion of PE-*v* samples with varying crosslink density (*i.e.*, PE-*v*-12-0.5, PE-*v*-12-0.3, PE-*v*-12-0.2). (A) Amplitude factor for scattering from a mass fractal and (B) fractal dimension. Error bars represent the standard error of the nonlinear regression.